\documentclass[longauth]{aa}

\usepackage{txfonts}
\usepackage{natbib}
\usepackage{graphicx}
\bibpunct{(}{)}{;}{a}{}{,} 

\begin{document}


\title{TNOs are Cool: A survey of the trans-Neptunian region V. \\ Physical characterization of 18 Plutinos using Herschel PACS observations\thanks{{\it Herschel} is an ESA space observatory with science instruments provided by European-led Principal Investigator consortia and with important participation from NASA. The Photodetector Array Camera and Spectrometer (PACS) is one of Herschel's instruments. PACS has been developed by a consortium of institutes led by MPE (Germany) and including UVIE (Austria); KU Leuven, CSL, IMEC (Belgium); CEA, LAM (France); MPIA (Germany); INAF-IFSI/OAA/OAP/OAT, LENS, SISSA (Italy); IAC (Spain). This development has been supported by the funding agencies BMVIT (Austria), ESA-PRODEX (Belgium), CEA/CNES (France), DLR (Germany), ASI/INAF (Italy), and CICYT/MCYT (Spain).}}
\titlerunning{TNOs are Cool: Physical characterization of 18 Plutinos using PACS observations}
\authorrunning{M. Mommert et al.}

\author{M. Mommert \inst{\ref{inst_DLR}}
   \and A. W. Harris \inst{\ref{inst_DLR}}
   \and C. Kiss \inst{\ref{inst_Konkoly}}
   \and A. P\'{a}l \inst{\ref{inst_Konkoly}}
   \and P. Santos-Sanz \inst{\ref{inst_Meudon}}
   \and J. Stansberry \inst{\ref{inst_UoA}}
   \and A. Delsanti \inst{\ref{inst_Meudon}, \ref{inst_Marseille}}
   \and E. Vilenius \inst{\ref{inst_MPE}}
   \and T. G. M\"uller \inst{\ref{inst_MPE}}
   \and N. Peixinho \inst{\ref{inst_Nuno1}, \ref{inst_Nuno2}} 
   \and E. Lellouch \inst{\ref{inst_Meudon}}
   \and N. Szalai \inst{\ref{inst_Konkoly}}
   \and F. Henry \inst{\ref{inst_Meudon}}   
   \and R. Duffard \inst{\ref{inst_Granada}}
   \and S. Fornasier \inst{\ref{inst_Meudon}, \ref{inst_Meudon2}}
   \and P. Hartogh \inst{\ref{inst_MPS}}
   \and M. Mueller \inst{\ref{inst_migo1}, \ref{inst_migo2}}
   \and J. L. Ortiz \inst{\ref{inst_Granada}}
   \and S. Protopapa \inst{\ref{inst_MPS}, \ref{inst_maryland}}
   \and M. Rengel \inst{\ref{inst_MPS}}
   \and A. Thirouin \inst{\ref{inst_Granada}}
  }

\institute{Deutsches Zentrum f\"ur Luft- und Raumfahrt (DLR), Institut f\"ur Planetenforschung, Rutherfordstra\ss e 2, 12489 Berlin, Germany; email: \texttt{michael.mommert@dlr.de} \label{inst_DLR}
      \and Max-Planck-Institut f\"ur extraterrestrische Physik (MPE), Postfach 1312, Giessenbachstr., 85741 Garching, Germany \label{inst_MPE}
      \and Konkoly Observatory of the Hungarian Academy of Sciences, H-1525 Budapest, P.O. Box 67, Hungary \label{inst_Konkoly}
      \and LESIA-Observatoire de Paris, CNRS, UPMC Univ Paris 06, Univ. Paris-Diderot, 5 Place J. Janssen, 92195 Meudon Pricipal Cedex, France \label{inst_Meudon}
      \and The University of Arizona, Tucson AZ 85721, USA \label{inst_UoA}
      \and Center for Geophysics of the University of Coimbra, Av. Dr. Dias da Silva, 3000-134 Coimbra, Portugal \label{inst_Nuno1}
      \and Astronomical Observatory of the University of Coimbra, Almas de Freire, 3040-04 Coimbra, Portugal \label{inst_Nuno2}
      \and Max-Planck-Institut f\"ur Sonnensystemforschung (MPS), Max-Planck-Stra\ss e 2, 37191 Katlenburg-Lindau, Germany \label{inst_MPS}
      \and SRON Netherlands Institute for Space Research, Postbus 800, 9700 AV
      Groningen, The Netherlands \label{inst_migo1}
      \and UNS-CNRS-Observatoire de la C\^ote d'Azur, Laboratoire Cassiop\'ee,
      BP 4229, 06304 Nice Cedex 04, France \label{inst_migo2}
      \and Univ. Paris Diderot, Sorbonne Paris Cit\'{e}, 4 rue Elsa Morante, 75205 Paris, France \label{inst_Meudon2} 
      \and Laboratoire d'Astrophysique de Marseille, CNRS \& Universit\'{e} de Provence, 38 rue Fr\'{e}d\'{e}ric Joliot-Curie, 13388 Marseille cedex 13, France \label{inst_Marseille}
      \and Instituto de Astrof\'{\i}sica de Andaluc\'{\i}a (CSIC) C/ Bajo de Hu\'{e}tor, 50, 18008 Granada, Spain \label{inst_Granada}
      \and Department of Astronomy, University of Maryland, College Park, MD
      20742 \label{inst_maryland}
     } 

\date{Received December 2, 2011 / Accepted January 31, 2012}

\abstract{The Herschel Open Time Key Programme \textit{TNOs are Cool: A survey of the trans-Neptunian region} aims to derive physical and thermal properties for a set of $\sim$140 Centaurs and Trans-Neptunian Objects (TNOs), including resonant, classical, detached and scattered disk objects. One goal of the project is to determine albedo and size distributions for specific classes and the overall population of TNOs.}  
         {We present Herschel PACS photometry of 18 Plutinos and determine sizes and albedos for these objects using thermal modeling. We analyze our results for correlations, draw conclusions on the Plutino size distribution, and compare to earlier results.} 
	 {Flux densities are derived from PACS mini scan-maps using specialized data reduction and photometry methods. In order to improve the quality of our results, we combine our PACS data with existing Spitzer MIPS data where possible, and refine existing absolute magnitudes for the targets. The physical characterization of our sample is done using a thermal model. Uncertainties of the physical parameters are derived using customized Monte Carlo methods. The correlation analysis is performed using a bootstrap Spearman rank analysis.} 
	 {We find the sizes of our Plutinos to range from $150$ to $730$~km and geometric albedos to vary between $0.04$ and $0.28$. The average albedo of the sample is $0.08\pm0.03$, which is comparable to the mean albedo of Centaurs, Jupiter Family comets and other Trans-Neptunian Objects. We were able to calibrate the Plutino size scale for the first time and find the cumulative Plutino size distribution to be best fit using a cumulative power law with $q=2$ at sizes ranging from 120--400~km and $q=3$ at larger sizes. We revise the bulk density of 1999~TC36 and find $\varrho=0.64_{-0.11}^{+0.15}\,\mbox{g cm}^{-3}$. On the basis of a modified Spearman rank analysis technique our Plutino sample appears to be biased with respect to object size but unbiased with respect to albedo. Furthermore, we find biases based on geometrical aspects and color in our sample. There is qualitative evidence that icy Plutinos have higher albedos than the average of the sample.} 
	 {} 


\maketitle


\section{Introduction}

Since its discovery in 1930, Pluto has been a unique object not only for being the only rocky planet-sized object not bound to a planet outside the orbit of Mars, but also for having the most eccentric and inclined orbit of all the planets, which even overlaps the orbit of Neptune. \citet{Cohen1965} were the first to show that, despite that overlap, close approaches between the planets are prevented by the 2:3 mean motion resonance: Pluto's revolution period equals 3/2 of Neptune's period, ensuring that conjunctions always occur near Pluto's aphelion. This leads to a high degree of stability of the orbit. The origin of Pluto's peculiar orbit ($e=0.25$ and $i=17\degr$) was first explained by \citet{Malhotra1993}, who showed that encounters of the Jovian planets with residual planetesimals during the late stages of the formation of the Solar System could lead to a radial migration of the former. As a result of Neptune's outward migration, a Pluto-like body could have been captured in the 2:3 resonance, excited to its highly inclined and eccentric orbit and transported outward.

In the meantime, the discovery of 1992 QB1 by \citet{Jewitt1993} showed that Pluto is not the only object beyond the orbit of Neptune. The following years revealed a large population of trans-Neptunian objects (TNOs), presumably a population of residual planetesimals from the age of the formation of the Solar System, as previously proposed by \citet{Edgeworth1949} and \citet{Kuiper1951}. The population shows dynamical complexity and recurring orbital characteristics allowing the classification of TNOs in different dynamical groups \citep[cf. ][]{Elliot2005, Gladman2008}. Many of the newly discovered objects show a dynamical behaviour similar to that of Pluto, which led \citet{Jewitt1996} to dub them \textit{Plutinos}. 

Plutinos reside in Neptune's 2:3 resonance \citep{Gladman2008} and are the most numerous resonant population. This observation agrees with dynamical studies by \citet{Melita2000}, which reveals that the 2:3 resonance is much more stable than other resonances, particularly at low inclinations. While the semi-major axes of Plutinos are strongly concentrated around 39.5 AU, their eccentricities and inclinations vary significantly from Pluto's ($0.03 \leq e\leq0.88$ and $0.4\degr\leq i\leq 40.2\degr$ with a mean eccentricity and inclination of 0.21 and 12.1$\degr$, respectively)\footnote[1]{Statistical data are based on Minor Planet Center data as of 18 Nov. 2011 (http://minorplanetcenter.net/iau/MPCORB.html).}. The origin of the Plutino population is ascribed to the same resonance capture mechanism which is responsible for Pluto's peculiar orbit \citep{Malhotra1995}. Based on numerical simulations, \citet{Duncan1995} find that the 2:3 mean motion resonance interferes with secular resonances leading to a severe instability for high inclination orbits and a longterm leakage for low inclination orbits of Plutinos, making them a possible source region of Jupiter Family Comets (JFCs). \citet{Morbidelli1997} found that a slow chaotic diffusion of Plutinos can be provided, which is necessary to explain the observed continuous flux of JFCs. Their results were recently confirmed by \citet{DiSisto2010}, who found that Plutinos may also be a secondary source of the Centaur population. A collisional analysis of the 2:3 resonance carried out by \citet{deElia2008} shows that the Plutino population larger than a few kilometers in diameter is not significantly altered by catastrophic collisions over the age of the Solar System. They also pointed out the importance of specifying the number of Pluto-sized objects among the Plutino population, since the escape frequency of Plutinos strongly depends on this number. 

\begin{figure}[b]
 \centering
 \resizebox{\hsize}{!}{\includegraphics{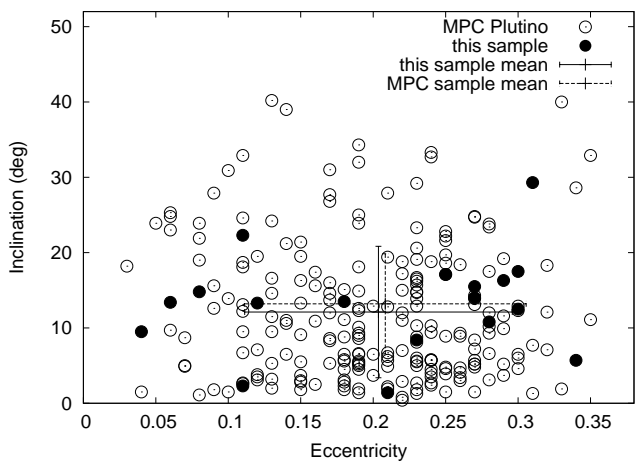}}
 \caption{Our Plutino sample (filled circles) compared to the known Plutino population (open circles, data from MPC, as of 18 Nov. 2011) in $e$--$i$ phase space. The crosses with solid and dashed error bars refer to the mean and the standard deviation of our sample and the known Plutino population, respectively. This plot demonstrates that our sample is heterogenous and reflects the dynamical spread of the complete Plutino population. (The plot omits 3 known Plutinos, due to their high eccentricities, which are not part of our sample; however, these objects are included in the determination of the mean.). }
 \label{fig:sample}
\end{figure}

Our sample of 18 Plutinos is selected based on the object classification
scheme of \citet{Gladman2008} and represents $\sim 7.5$\% of the known Plutino
population as of 27 October, 2011 (data from MPC). 14 of our 18 sample targets have been classified in \citet{Gladman2008}. For the remaining ones dynamical simulations performed by Ch. Ejeta show that their Plutino-type orbits are stable for at least $10^7$~years, which suggests, but does not prove, orbital stability over the lifetime of the Solar System. We give the numbers and preliminary designations of the sample objects in Table \ref{tab:sample}. In order to avoid confusion, we will refer to specific objects in this work only by their preliminary designation or name, where available. In Figure \ref{fig:sample} we plot our dataset in comparison to the sample of all known Plutinos in $e$--$i$ phase space. We neglect $a$ here, since it is very well confined around 39.5~AU. The plot shows the spread of both samples in $e$--$i$ space, whereas the means of the two samples show a good agreement. Therefore, we assume that our sample well represents the dynamical variety of the whole Plutino population.

Our Plutino sample is selected solely on the basis of Herschel observability in the sense of observation geometry and detector sensitivity. Herschel observations presented here are conducted over a time span of more than 7 months. In this time span, Herschel is able to observe at all ecliptical longitudes, despite its restrictions on the solar aspect angle, which must not be lower than 60.8$\degr$ and not be larger than 119.2$\degr$. Hence, no bias is introduced based on preferred ecliptical longitudes. Since the target objects of the `TNOs are Cool' project are selected on the basis of optical discoveries, the most important bias in our sample is presumably the optical detection bias, favoring large objects and objects with high albedos. The detectional bias impact on our results will be discussed in Section \ref{ref:discussion_correlation}.

Our target sample includes 3 known multiple systems, which are Pluto, 1999~TC36 and 2003~AZ84 and the expected binary 2002~GN171. We emphasize, that the spatial resolution of PACS is not sufficient to separate any multiple system. Flux densities measured for one object therefore consist of contributions from fluxes of all system components. Diameter and albedo determined with our models refer to respective parameters of an object with the area equivalent diameter of the whole system. We discuss these objects in detail in Section \ref{ref:discussion}.

The `TNOs are Cool: A Survey of the Trans-Neptunian Region' project is a Herschel Open Time Key Programme awarded some 370 h of Herschel observing time for investigating about 140 trans-Neptunian objects with known orbits \citep{Mueller2009}. The observations include PACS \citep{Poglitsch2010} and SPIRE \citep{Griffin2010} point-source photometry. The goal is to characterize the diameter and albedo for the individual objects and the full sample using radiometric techniques, in order to probe formation and evolution processes in the Solar System. Using thermal modeling we determine sizes and albedos of the 18 Plutino members. The `TNOs are Cool' sample includes data of 7 more Plutinos, which have not been processed by the time of writing this.


\begin{table}[t]
 \caption{Plutino target sample. We list numbers and preliminary designations or object names, where allocated. In order to avoid confusion, we will refer to specific objects in this work only by their preliminary designation or name. }
 \centering
 \begin{tabular}{rllrl}
 \hline \\ [-7pt]
 (15875)  & 1996~TP66  & ~ & (144897) & 2004~UX10  \\ 
 (38628)  & 2000~EB173~Huya       & ~ & (175113) & 2004~PF115 \\
 (47171)  & 1999~TC36  & ~ & (208996) & 2003~AZ84  \\ 
 (47932)  & 2000~GN171 & ~ &        ~ & 2001~KD77  \\ 
 (55638)  & 2002~VE95  & ~ &        ~ & 2001~QF298 \\ 
 (84719)  & 2002~VR128 & ~ &        ~ & 2002~VU130 \\
 (84922)  & 2003~VS2   & ~ &        ~ & 2002~XV93  \\
 (120216) & 2004~EW95  & ~ &        ~ & 2003~UT292 \\ 
 (134340) & Pluto      & ~ &        ~ & 2006~HJ123 \\
 \hline
 \end{tabular}
 \label{tab:sample}
\end{table}


\begin{table*}[!t]
 \caption{PACS observations summary. Column headings are: object name; Herschel ObsID of the first AOR of a sequence of four AORs of two consecutive observations (first visit/follow-on observation), respectively; total duration of all AORs in minutes; observation mid-time of all observations (MM-DD UT) in 2010; $r$: heliocentric distance; $\Delta$: Herschel-target distance; $\alpha$: solar phase angle; color-corrected flux density values at PACS photometer reference wavelengths 70, 100 and 160 $\mu m$, uncertainties include photometric and calibration uncertainties. Upper limits are 1~$\sigma$.}
 \centering
 \begin{tabular}{rccccccccc}
 \hline \hline \\ [-7pt]
 Object & ObsIDs & Dur. &  Mid-time & $r$    & $\Delta$ & $\alpha$ &  \multicolumn{3}{c}{PACS Flux Densities (mJy)} \\
        &       & (min)&   (UT)    &   (AU) & (AU)     & (\degr)  &   70~$\mu m$   &   100~$\mu m$   &   160~$\mu m$ \\
 \hline \\[-7pt]
 1996~TP66              & 1342202289/...2310 & 113.3 & 08-08~03:28 & 27.3175 & 27.6467 & 2.0 &  $\leq0.8$       & $\leq1.1$   & $\leq1.5$ \\ 
 1999~TC36              & 1342199491/...9630 & 75.7 & 07-01~10:10 & 30.6722 & 30.8956 & 1.9 &   $27.2\pm1.4$ & $22.3\pm1.9$  & $11.0\pm1.6$ \\ 
 2000~GN171             & 1342202906/...2971 & 150.9 & 08-12~18:19 & 28.2876 & 28.4320 & 2.0 &  $3.2\pm0.7$  & $5.8\pm1.1$   & $3.2\pm1.3$ \\ 
 2001~KD77              & 1342205966/...6009 & 150.9 & 10-07~05:54 & 35.7854 & 36.1111 & 1.5 &  $5.4\pm0.6$  & $1.0\pm1.1$   & $4.1\pm1.8$ \\ 
 2001~QF298             & 1342197661/...7681 & 113.3 & 06-03~06:43 & 43.1037 & 43.3727 & 1.3 &  $7.2\pm0.8$  & $6.5\pm1.7$  & $5.0\pm1.3$ \\
 2002~VE95              & 1342202901/...2953 & 113.3 & 08-12~15:17 & 28.5372 & 28.8990 & 1.9 &  $10.6\pm0.8$ & $8.6\pm1.1$  & $6.8\pm1.6$ \\
 2002~VR128             & 1342190929/...0990 & 109.3 & 02-22~11:57 & 37.4636 & 37.7851 & 1.4 &  $15.8\pm0.9$ & $13.1\pm1.2$ & $8.8\pm1.3$ \\ 
 2002~VU130             & 1342192762/...2783 & 112.8 & 03-26~05:04 & 41.6877 & 42.1392 & 1.2 &  $3.2\pm0.8$  & $2.4\pm1.0$  & $2.1\pm1.3$ \\
 2002~XV93              & 1342193126/...3175 & 112.8 & 03-31~22:35 & 39.7152 & 40.0645 & 1.4 &  $17.3\pm1.1$ & $17.4\pm1.2$ & $10.8\pm2.1$ \\
 2003~AZ84$^{a}$        & 1342187054         & \multicolumn{2}{c}{cf. \citep{Mueller2010}} & 45.376 & 44.889 & 1.1 & $27.0\pm2.8$ & - & $19.7\pm5.2$ \\
 2003~AZ84$^{b}$        & cf. Footnote {\it b} & 484.5 & 09-27~13:42 & 45.3025 & 45.6666 & 1.2 &       --      & $25.7\pm0.3$ & $14.6\pm0.8$\\
 2003~UT292             & 1342190949/...1025 & 145.6 & 02-22~22:33 & 29.4217 & 29.4484 & 1.9 &  $6.3\pm0.8$  & $4.7\pm1.4$  & $3.6\pm1.2$ \\
 2003~VS2               & 1342191937/...1977 & 75.7  & 03-10~07:22 & 36.4694 & 36.7093 & 1.5 &  $17.8\pm1.1$ & $16.5\pm1.5$ & $10.2\pm3.0$ \\ 
 2003~VS2$^{c}$         & cf. Footnote {\it c} & 508.0 & 08-11~18:23 & 36.4760 & 36.8017 & 1.5 &  $14.4\pm0.3$ &      --      & $14.0\pm0.6$ \\
 2004~EW95              & 1342199483/...9712 & 113.3 & 07-01~22:42 & 27.4708 & 27.1723 & 2.1 &  $19.5\pm0.9$ & $18.7\pm1.2$ & $9.6\pm1.8$ \\ 
 2004~PF115             & 1342208462/...8841 & 113.3 & 11-11~10:53 & 41.4271 & 41.2712 & 1.4 &  $10.7\pm0.9$ & $10.6\pm1.0$ & $8.8\pm2.1$ \\
 2004~UX10              & 1342199495/...9626 & 75.7  & 07-01~11:12 & 38.9500 & 39.3307 & 1.4 &  $8.7\pm1.1$  & $10.9\pm1.6$  & $5.2\pm1.8$ \\
 2006~HJ123             & 1342204150/...4200 & 113.3 & 09-09~05:52 & 36.5383 & 36.9867 & 1.4 &  $3.0\pm1.2$  & $3.5\pm1.6$  & $3.2\pm2.1$  \\ 
 Huya                   & 1342202873/...2914 & 75.7  & 08-12~03:37 & 28.6648 & 28.7665 & 2.0 &  $41.4\pm1.6$ & $37.6\pm1.8$ & $22.5\pm2.2$ \\ 
 Pluto/Charon & 1342191953/...1988 & 75.7  & 03-10~12:46 & 31.7985 & 32.0470 & 1.7 & $283.9\pm8.6$ & $354.8\pm11.2$ & $289.2\pm17.2$\\ 
 \hline
 \end{tabular}
 \tablefoot{Geometric data were extracted from NASA Horizons (http://ssd.jpl.nasa.gov/horizons.cgi) for the indicated mid-time of the observations; 
  \tablefoottext{a}{chop/nod observations \citep{Mueller2010};}
  \tablefoottext{b}{averaged lightcurve observations using ObsIDs 1342205152/...5222-5225 sampling the 100 and 160~$\mu m$ bands;}
  \tablefoottext{c}{averaged lightcurve observations using ObsIDs 1342202371/...2574-2577 sampling the 70 and 160~$\mu m$ bands.}
 }
 \label{tab:PACS_observations}
\end{table*}

\begin{table*}[!t]
 \caption{Absolute magnitude $H$, lightcurve information, information on the presence of ices and color indices. The top part of the table shows: object name; $\alpha$: phase angle range of observations utilized in the determination of $H$; $N$: number of available observations; $\beta$: linear phase coefficient; $H_V$: absolute magnitude in the $V$ band with respective uncertainty; photometry references; $\Delta_{mag}$: optical lightcurve amplitude (peak-to-peak); Ice?: information on the presence of ices, if available (discoveries in parentheses are tentative); LC \& Ice Ref.: lightcurve and ice references. The lower part of the table shows different color indices for our sample objects, where available, including uncertainties, and the spectral slope $s$ in percent of reddening per 100~$nm$ as determined from the given color information following \citet{HainautMBOSS}, and references. The weighted sample mean (cf. Table \ref{tab:optical_photometry}) of each color index is given for comparison, as well as some color indices of the Sun. }
 \centering
 \begin{tabular}{rcccccccl}
 \hline \hline \\ [-7pt]
 Object       & $\alpha$    & $N$  & $\beta$           & $H_V$            & Photometry    & $\Delta_{mag}$    & Ice?  & LC \& Ice \\
              & (\degr)     &      & (mag/\degr)       & (mag)            & References    &      (mag)        &       & Ref. \\ [+2pt]
 \hline \\ [-7pt]
 1996~TP66    & 0.90-1.86   &  3  & 0.10$\pm$0.04     & 7.51$\pm$0.09     & \multicolumn{1}{l}{1,2,3}         & $<$0.04           &  no   & 4,36  \\ 
 1999~TC36    & 0.28-1.71   & 47  & 0.08$\pm$0.04     & 5.41$\pm$0.10$^a$ & \multicolumn{1}{l}{1,8-13}        & 0.20$\pm$0.04     &  H$_2$O  & 12,16,36,41,42,43 \\ 
 2000~GN171   & 0.02-2.04   & 40  & 0.14$\pm$0.03$^c$ & 6.45$\pm$0.34$^a$ & \multicolumn{1}{l}{10,13,14}      & 0.61$\pm$0.03     &  no   & 17,36,39,40,42 \\ 
 2001~KD77    & 1.54-1.56   &  6  & 0.10$\pm$0.04     & 6.42$\pm$0.08     & \multicolumn{1}{l}{20,21}         & $<$0.07           &       & 22 \\ 
 2001~QF298   & 0.68-1.02   &  3  & 0.10$\pm$0.04     & 5.43$\pm$0.07     & \multicolumn{1}{l}{21,24,25}      & $<$0.12           &      & 22 \\
 2002~VE95    & 0.57-2.07   & 40  & 0.16$\pm$0.04     & 5.70$\pm$0.06     & \multicolumn{1}{l}{13,26}         & 0.08$\pm$0.04     &  H$_2$O  & 26,27,36 \\
 2002~VR128   & 0.31-0.58   &  6  & 0.10$\pm$0.04     & 5.58$\pm$0.37     & \multicolumn{1}{l}{MPC}           &       -           &       & -  \\ 
 2002~VU130   & 1.28-1.37   &  3  & 0.10$\pm$0.04     & 5.47$\pm$0.83     & \multicolumn{1}{l}{MPC}           &       -           &       & -  \\
 2002~XV93    & 1.11        &  3  & 0.10$\pm$0.04     & 5.42$\pm$0.46     & \multicolumn{1}{l}{MPC}           &       -           &       & -  \\
 2003~AZ84    & 0.33-1.17   &  4  & 0.15$\pm$0.05     & 3.74$\pm$0.08$^a$ & \multicolumn{1}{l}{14,24,28,29}   & 0.14$\pm$0.03     &  H$_2$O  & 27,36,42,48 \\ 
 2003~UT292   & 0.35-1.75   & 13  & 0.10$\pm$0.04     & 6.85$\pm$0.68$^b$ & \multicolumn{1}{l}{MPC}           &       -           &       & -  \\
 2003~VS2     & 0.53-0.59   &  7  & 0.10$\pm$0.04     & 4.11$\pm$0.38     & \multicolumn{1}{l}{MPC}           & 0.21$\pm$0.01     &  H$_2$O     & 30,36,44 \\ 
 2004~EW95    & 0.45-0.80   & 20  & 0.10$\pm$0.04     & 6.69$\pm$0.35     & \multicolumn{1}{l}{MPC}           &       -           &       & -  \\ 
 2004~PF115   & 0.31-0.93   & 13  & $0.10\pm0.04$     & $4.54\pm0.25^b$   & \multicolumn{1}{l}{MPC}           & -                 &       & -  \\
 2004~UX10    & 0.75-1.06   &  4  & $0.10\pm0.04$     & $4.75\pm0.16$     & \multicolumn{1}{l}{28,30}         & $0.08\pm0.01$     &   (H$_2$O) & 30,48 \\
 2006~HJ123   & 0.22-1.49   &  5  & 0.10$\pm$0.04     & 5.32$\pm$0.66$^b$ & \multicolumn{1}{l}{MPC}           &       -           &       & -  \\
 Huya         & 0.49-1.80   & 13  & 0.09$\pm$0.04     & 5.14$\pm$0.07     & \multicolumn{1}{l}{8,10,32}       & $<$0.1            &   (H$_2$O) & 16,36,37,38,39,40 \\ 
 Pluto/Charon & 0.18-1.87   & 277 & 0.037$\pm$0.002$^d$ & $-0.67\pm0.34$  & \multicolumn{1}{l}{MPC}           & 0.33              &   CH$_4$,CO,N$_2$    & 33,45,46,47 \\ [+3pt] 
 \hline \\ [-7pt]
 Object      & B-V             & V-R             & B-R             & R-I             & V-I             & B-I             & s         & Ref.\\ [+2pt]
 \hline \\ [-7pt]
 1996~TP66   & 1.03$\pm$0.11 & 0.66$\pm$0.07 & 1.68$\pm$0.12 & 0.66$\pm$0.10 & 1.29$\pm$0.11 & 2.25$\pm$0.15 & 30.4$\pm$4.7 & 1,3,5,6,7,21 \\ 
 1999~TC36   & 1.00$\pm$0.13 & 0.70$\pm$0.03 & 1.74$\pm$0.05 & 0.62$\pm$0.05 & 1.30$\pm$0.13 & 2.35$\pm$0.14 & 32.1$\pm$2.3 & 1,8,9,11-13,15\\ 
 2000~GN171  & 0.96$\pm$0.06 & 0.60$\pm$0.04 & 1.56$\pm$0.07 & 0.62$\pm$0.05 & 1.14$\pm$0.17 & 2.16$\pm$0.09 & 23.9$\pm$2.9 & 13,14,18,19\\ 
 2001~KD77   & 1.12$\pm$0.05 & 0.62$\pm$0.07 & 1.76$\pm$0.06 & 0.57$\pm$0.07 & 1.20$\pm$0.07 & 2.31$\pm$0.09 & 23.5$\pm$4.1 & 20,21,23\\ 
 2001~QF298  & 0.67$\pm$0.07 & 0.39$\pm$0.06 & 1.05$\pm$0.09 & 0.57$\pm$0.19 & 0.89$\pm$0.19 & 1.35$\pm$0.09 & 4.6$\pm$4.6  & 19,21,24,25\\
 2002~VE95   & 1.07$\pm$0.14 & 0.72$\pm$0.05 & 1.79$\pm$0.04 & 0.76$\pm$0.12 & 1.38$\pm$0.15 & 2.47$\pm$0.13 & 37.8$\pm$3.7 & 13,19,28\\
 2002~VR128  & 0.94$\pm$0.03 & 0.60$\pm$0.02 & 1.54$\pm$0.04 &        -      &        -      &      -        & 22.76$\pm$2.1 & 19\\ 
 2002~XV93   & 0.72$\pm$0.02 & 0.37$\pm$0.02 & 1.09$\pm$0.03 &        -      &        -      &      -        & 0.9$\pm$2.1 & 19\\
 2003~AZ84   & 0.67$\pm$0.05 & 0.38$\pm$0.04 & 1.05$\pm$0.06 & 0.55$\pm$0.15 & 0.92$\pm$0.14 & 1.68$\pm$0.19 & 3.6$\pm$3.5 & 19,24,28,29,34\\ 
 2003~VS2    & 0.93$\pm$0.02 & 0.59$\pm$0.02 & 1.52$\pm$0.03 &        -      &        -      &      -        & 21.7$\pm$2.1 & 19\\ 
 2004~EW95   & 0.70$\pm$0.02 & 0.38$\pm$0.02 & 1.08$\pm$0.03 &        -      &        -      &      -        & 1.7$\pm$2.1 & 19\\ 
 2004~UX10   & 0.95$\pm$0.02 & 0.58$\pm$0.05 & 1.53$\pm$0.02 &        -      &        -      &      -        & 20.2$\pm$4.4 & 28,31\\
 Huya        & 0.95$\pm$0.05 & 0.57$\pm$0.09 & 1.54$\pm$0.06 & 0.61$\pm$0.05 & 1.19$\pm$0.06 & 2.14$\pm$0.07 & 21.9$\pm$4.6 & 7,8,10,18,32\\ [+3pt]
 Sample av.  & 0.84$\pm$0.13 & 0.51$\pm$0.12 & 1.43$\pm$0.25 & 0.62$\pm$0.05 & 1.19$\pm$0.11 & 2.07$\pm$0.35 & 17$\pm$12 & \\ 
 Sun         & 0.64          & 0.36          &       -       &        -      & 0.69          &      -        &    - &  35\\ [+3pt]
 \hline
 \end{tabular}
 \tablefoot{We determine the weighted mean $\langle x \rangle$ of quantities $x_i$ using the absolute uncertainties $\sigma_i$ as weighting parameter. The uncertainty of $\langle x \rangle$, $\sigma$, is calculated as $\sigma^2 = \left(1+\sum_i[(x_i-\langle x \rangle)^2/\sigma_i^2]\right)/\sum_i (1/\sigma_i^2).$ Defined this way, $\sigma$ is a combination of the weighted root mean square and the weighted standard deviation of the uncertainties.
   \tablefoottext{a}{original photometric uncertainty is smaller than the half peak-to-peak lightcurve amplitude: new value is $(\sigma_H^2 + (0.5\Delta_{mag})^2)^{1/2}$};
   \tablefoottext{b}{converted $R$ band data};
   \tablefoottext{c}{regression analysis leads to an unrealistic phase coefficient, adopt instead phase coefficient from \citet{Belskaya2008}};
   \tablefoottext{d}{phase coefficient adopted from \citet{Buie1997}};
  \textbf{References.} MPC: photometric data were provided by the Minor Planet Center observations database (http://minorplanetcenter.net/iau/ECS/MPCOBS/MPCOBS.html);
    (1):  \citet{Boehnhardt2001};
    (2):  \citet{Davies2000};
    (3):  \citet{Jewitt1998};
    (4):  \citet{Collander-Brown1999};
    (5):  \citet{Tegler1998};
    (6):  \citet{Barucci1999};
    (7):  \citet{Jewitt2001};
    (8):  \citet{Doressoundiram2001};
    (9):  \citet{Delsanti2001};
    (10): \citet{McBride2003};
    (11): \citet{Tegler2003};
    (12): \citet{Dotto2003};
    (13): \citet{Rabinowitz2007};
    (14): \citet{DeMeo2009};
    (15): \citet{Benecchi2009};
    (16): \citet{Ortiz2003};
    (17): \citet{Sheppard2002};
    (18): \citet{Boehnhardt2001};
    (19): Tegler, private communication; 
    (20): \citet{Doressoundiram2002};
    (21): \citet{Doressoundiram2007};
    (22): \citet{Sheppard2003};
    (23): \citet{Peixinho2004};
    (24): \citet{Fornasier2004};
    (25): \citet{Doressoundiram2005};
    (26): \citet{Barucci2006};
    (27): \citet{Ortiz2006};
    (28): \citet{Perna2010};
    (29): \citet{SantosSanz2009};
    (30): \citet{Thirouin2010};
    (31): \citet{Romanishin2010};
    (32): \citet{Ferrin2001};
    (33): \citet{Buie1997};
    (34): \citet{Rabinowitz2008};
    (35): \citet{Doressoundiram2008};
    (36): \citet{Barkume2008};
    (37): \citet{Licandro2001};
    (38): \citet{Brown2000};
    (39): \citet{deBergh2004};
    (40): \citet{Alvarez-Candal2007};
    (41): \citet{Merlin2005};
    (42): \citet{Guilbert2009};
    (43): \citet{Protopapa2009};
    (44): \citet{Barucci2011VE95};
    (45): \citet{Owen1993};
    (46): \citet{DeMeo2010};
    (47): \citet{Merlin2010};
    (48): \citet{Barucci2011}.
 }
 \label{tab:optical_photometry}
\end{table*}

\section{Observations and Data Reduction}

\subsection{Herschel Observations}

Photometric measurements using the Photometer Array Camera and Spectrometer \citep[PACS, ][]{Poglitsch2010} onboard the Herschel Space Observatory \citep{Pilbratt2010} have been taken in mini scan map mode covering homogeneously a field of roughly 1\arcmin\, in diameter. This mode turned out to be best suited for our needs and offers more sensitivity than other observation modes \citep{Mueller2010}. 

According to the `TNOs are Cool!' Open Time Key Programme observation strategy a target is observed at two epochs, separated by a time interval that corresponds to a movement of 30--50\arcsec\, of the target, allowing for an optimal background subtraction, to eliminate confusion noise and background sources. At each epoch the target is observed in the `blue' (nominal wavelength of 70~$\mu m$) and `green' (100~$\mu m$) band twice, using two different scan position angles. `Red' (160~$\mu m$) band data are taken in parallel when sampling one of the other bands. This forms a series of 4 measurements in the blue and green bands, and a series of 8 measurements in the red band for a specific 
target. The maps are taken in the medium scan speed (20\arcsec/sec) mode, using a scan-leg length of 3\arcmin and 2--4 repetitions. In the case of 2002~VR128 a scan-leg length of 2.5\arcmin\, was used. More details on the observation planning can be found in \citet{Vilenius2011}.

Observational circumstances and PACS flux densities are summarized in Table \ref{tab:PACS_observations}. Additional information on the targets is given in Table \ref{tab:optical_photometry}.

\subsection{Herschel PACS Data Processing}
\label{ref:data_reduction}

The raw PACS measurements were used to produce individual scan maps (level-2) using an optimized version of the standard PACS mini scan map pipeline in HIPE \citep{Ott2010}. The individual scan maps of the same epoch and band were mosaicked using the MosaicTask(). In the production of the final maps background-matching and source stacking techniques are applied to correct for the possible relative astrometric uncertainties between the two visits. We created two sorts of background-eliminated products from the `per-visit' mosaics: `background-subtracted' \citep[see][]{Stansberry2008,SantosSanz2011} and `double-differential' maps\footnote[2]{Background-subtracted mosaics were used in the cases of 1996~TP66, 2001~KD77, 2002~VE95, 2002~XV93, 2003~AZ84, 2003~VS2, 2004~PF115 and Pluto/Charon; double-differential mosaics for the other targets.}. The latter are generated by subtracting the maps of the two visits, yielding a positive and a negative beam of the object on the differential map with all background structures eliminated. A duplicate of this map is then shifted in such a way as to match the positive beam of the original map with the negative one of the duplicate map. In the last step the original and shifted mosaics are subtracted again and averaged, resulting in a double-differential mosaic with a positive beam showing the full target flux, and two negative beams on either side showing half the target flux. The final photometry is performed on the central, positive beam. The advantage of this technique is the nearly complete elimination of the sky background that makes it favorable for faint targets. Full details can be found in \citet{Kiss2012}.

Photometry is performed on the final background-subtracted and/or double-differential maps, which are both background-eliminated. Flux densities are derived via aperture photometry using either IRAF/Daophot or IDL/Astrolib routines, both producing identical results. We extract the flux at the photocenter position of the target and apply an aperture correction technique \citep{Howell1989} for each aperture radius based on the encircled energy fraction for the PACS point spread function \citep{Mueller2011PACScalibration}. We construct an aperture-corrected curve of growth from which we derive the optimum synthetic aperture, which usually lies in the 'plateau' region of the curve of growth.

Photometric uncertainties are estimated by random implantation of 200 artificial sources in the nearby sky background of the target. The 1\,$\sigma$ photometric uncertainty of the target flux is derived as the standard deviation of these artificial source fluxes \citep{SantosSanz2011}. This 1\,$\sigma$ limit is also given as an upper limit in the case of non-detections. 

The absolute calibration of our data is based on mini scan maps of 5 fiducial stars and 18 large main belt asteroids \citep{Mueller2011PACScalibration}. The absolute calibration uncertainties of the PACS bands are 3\% for the 70 and 100~$\mu m$ bands and 5\% for the 160~$\mu m$ band, respectively. This additional uncertainty adds quadratically to the photometric error and is included in the flux densities given in Table \ref{tab:PACS_observations}. The absolute accuracy of the photometry was checked against the predicted standard star flux densities. In the bright regime (above 30\,mJy) the relative accuracy was found to be $\sim$5\%, while in the faint regime (below 30\,mJy) the accuracy is driven by the uncertainties in the maps and is about or below 2\,mJy in all bands, on the individual maps. 

A general description of the PACS data reduction steps (including photometry) will be given in \citet{Kiss2012}. 

\begin{table*}[t]
 \caption{Spitzer observations summary. Column headings are: object name; Spitzer AORKEY identification; total duration of all AORs; observation mid-time of the 24 and 70 $\mu m$ measurements (20YY-MM-DD UT); $r$: heliocentric distance; $\Delta$: Spitzer-target distance; $\alpha$: solar phase angle; color-corrected flux density values at MIPS photometer reference wavelengths 23.68 and 71.42 $\mu m$, upper limits are $1\,\sigma$; references.}
 \centering
 \begin{tabular}{rccccccccl}
 \hline \hline \\ [-7pt]
 Object & AORKEY & Dur. & Mid-time & $r$    & $\Delta$ & $\alpha$ & \multicolumn{2}{c}{MIPS Flux Densities (mJy)} & Ref.\\
        &        & (min)& (UT)     & (AU)   & (AU)     & (\degr)  &   24 $\mu m$     &    70 $\mu m$ \\
 \hline \\ [-7pt]
 1996~TP66   & 8805632  & 38.4 & 04-01-23~03:31 & 26.4913 & 26.2500 & 2.1 & $0.689\pm0.038$ & $<5.87$ & 1\\ 
             & 12659456 & 40.8 & 05-09-24~15:54 & 26.6292 & 26.1132 & 1.9 & $0.426\pm0.029$ & $<2.30$ & 1\\
 1999~TC36   & 9039104  & 62.4 & 04-07-12~11:04 & 31.0977 & 30.9436 & 1.9 & $1.233\pm0.022$ & $25.30\pm2.53$ & 1\\
 2000~GN171  & 9027840  & 25.1 & 04-06-21~00:01 & 28.5040 & 28.0070 & 1.8 & $0.258\pm0.031$ & $5.60\pm4.00$ & 2\\ 
 2002~VE95   & 17766912/...7168 &  18.0 & 07-09-18~21:28 & 28.2291 & 28.2962 & 2.1 & $0.476\pm0.044$ & - & 3\\  
 2002~XV93   & 17768704/...8960 & 227.4 & 07-10-28~04:30 & 40.0092 & 39.6962 & 1.4 & $0.321\pm0.018$ & - & 3\\  
 2003~AZ84   & 10679040 & 28.9  & 06-03-30~09:22 & 45.6674 & 45.218 & 1.1 & $0.291\pm0.023$ & $17.8\pm2.66$ & 1\\ 
 2003~VS2    & 10680064 & 11.7 & 05-08-27~10:08 & 36.4298 & 36.5344 & 1.6 & $0.304\pm0.051$ & $25.7\pm7.34$ & 1\\ 
 Huya        & 8808192  & 21.3 & 04-01-27~09:40 & 29.3261 & 29.2503 & 1.9 & $3.630\pm0.052$ & $57.2\pm5.25$ & 1\\
             & 8937216  & 43.1 & 04-01-29~17:55 & 29.3252 & 29.2100 & 1.9 & $3.400\pm0.050$ & $52.9\pm1.86$ & 1\\ 
 \hline
 \end{tabular}
 \tablefoot{Geometric data were extracted from NASA Horizons (http://ssd.jpl.nasa.gov/horizons.cgi) for the indicated mid-time of the observations;
  \textbf{References.}
   \tablefoottext{1}{\citet{Stansberry2008}, uncertainties derived as flux density/SNR};
   \tablefoottext{2}{revised 70~$\mu m$ data from \citet{Stansberry2008}};
   \tablefoottext{3}{previously unpublished data}.
 }
 \label{tab:MIPS_observations}
\end{table*}

The flux densities given in Table \ref{tab:PACS_observations} have been color corrected and therefore are monochromatic. The measured flux density is determined by the response function of the PACS band filters convolved with the spectral energy distribution (SED) of the source \citep{Mueller2011PACScc}. The SED of TNOs resembles a black body spectrum of a certain temperature; the resulting color-correction factors depend weakly on the emission temperature. The black body temperature is approximated by the mean surface temperature, which is given by $2^{-0.25}T_{SS}$, where $T_{SS}$ is the subsolar temperature, defined in Equation \ref{eqn:T_SS}. We determine the color correction factors based on the mean surface temperature in an iterative process during the modeling. The mean color correction factors for all objects in this sample are $0.9810\pm0.0003$, $0.988\pm0.003$ and $1.015\pm0.005$ for the 70, 100 and 160~$\mu m$ band, respectively: the size of the color corrections are smaller than the uncertainty of the individual flux density measurements (cf. Table \ref{tab:PACS_observations}), which justifies the use of the mean surface temperature as an approximation of the best fit black body temperature.

We have obtained time-resolved lightcurve observations of 2003~AZ84 and 2003~VS2, covering 110 \% and 106\% of the respective lightcurve using 99 and 100 observations, respectively. A detailed analysis of both lightcurves will be subject of an upcoming publication \citep{SantosSanz2012}. In this work, we make use of the averaged fluxes, which are more precise than single measurements, due to the larger number of observations and the cancelling of lightcurve effects.

\subsection{Spitzer MIPS Observations}
\label{label:spitzer}

Whenever possible, the Herschel flux densities were combined with existing flux density measurements of the Multiband Imaging
Photometer for Spitzer (MIPS, \citet{Rieke2004}) onboard the Spitzer Space Telescope \citep{Werner2004} to improve the results. Useful data were obtained in the
MIPS 24 and 70~$\mu m$ bands, which have effective wavelengths of 23.68 and 71.42~$\mu m$. The MIPS 70~$\mu m$ band overlaps with the Herschel 70~$\mu m$ channel, providing a consistency check between data from the two observatories. A comparison of the PACS and MIPS 70~$\mu m$ band flux densities shows that most measurements are consistent within a $1~\sigma$ range. Significant deviations occur for 2003~AZ84 and Huya, which might have been caused by lightcurve effects or statistical noise. 

The reduction of MIPS observations of TNOs has been described in detail in \citet{Stansberry2008} and \citet{Brucker2009}, details on the calibration can be found in \citet{Gordon2007} and \citet{Engelbracht2007}. We adopt absolute calibration uncertainties of 3\% and 6\% for the 24 and 70~$\mu m$ observations (50\% larger than the uncertainties derived for observations of stellar calibrators), respectively. The larger uncertainties account for effects from the sky-subtraction technique, the faintness of the targets, and uncertainties in the color corrections.

The sky subtraction techniques introduced in Section \ref{ref:data_reduction} are derived from techniques originally developed for the MIPS data reductions. Reprocessed fluxes presented here are based on new reductions of the MIPS data, utilizing updated ephemeris positions for the targets. This allows for more precise masking of the target when generating the image of the sky background, and for more precise placement of the photometric aperture. In most cases the new fluxes are very similar to the previously
published values for any given target, but in a few cases significant improvements in the measured flux density and SNR were achieved.

The standard color correction routine \citep{Stansberry2007} resulting in monochromatic flux densities for the 24 and 70~$\mu m$ bands requires the measurement of both bands and assumes the temperature of a black body which fits the 24:70 flux density ratio best. However, the two previously unpublished flux densities of 2002~VE95 and 2002~XV93 are for the 24~$\mu m$ band only, which precludes the application of the standard method. In order to provide approximate color correction to these flux densities, we apply a method similar to the Herschel PACS color correction routine.

\subsection{Optical Photometry}

In order to derive diameter and albedo estimates, we combine thermal infrared measurements with optical data in the form of the absolute magnitude $H$, which is the object's magnitude if it was observed at 1~$AU$ heliocentric distance and 1~$AU$ distance from the observer at a phase angle $\alpha=0$. $H$, albedo $p_V$ and diameter $d$ are related via 
\begin{equation}
d = 2 \delta \cdot 10^{V_{\sun}/5} \cdot 10^{-H/5}/\sqrt{p_V},  \label{eqn:HdpV}
\end{equation}
with the magnitude of the Sun $V_{\sun} = -26.76\pm0.02$ \citep{Bessell1998} in the Johnson-Cousins-Glass system and $\delta$ a numerical constant. The relation returns $d$ in km, if $\delta$ equals 1~AU expressed in km. In order to derive reliable albedo estimates, $H$ and its uncertainty have to be well known \citep{Harris1997}. This is not the case for most small bodies in the Solar System, including Plutinos, for which reliable data are usually sparse. We determine $H$ magnitudes from observed magnitudes in literature and observational data from the Minor Planet Center (MPC) (cf. Table \ref{tab:optical_photometry}). We deliberately do not use the $H$ magnitudes provided by the MPC, since there is no uncertainty estimate given for these values and the reliability of these magnitudes is questionable (for instance, \citet{Romanishin2005} have determined a systematic uncertainty of MPC $H$ magnitudes of 0.3 magnitudes). Our results include uncertainty assessments and are reasonably close other estimates (for instance \citet{Doressoundiram2007}).

$H$ and the geometric albedo are defined at zero phase angle ($\alpha$). However, existing photometric data were taken at $\alpha \neq 0$, which is corrected for by assuming a linear approximation of the phase angle dependence: $H_V=m_V(1,1)-\alpha \cdot \beta$, in which $m_V(1,1)$ is the apparent $V$ band magnitude normalized to unity heliocentric and geocentric distance, and $\beta$ is the linear phase coefficient. Since we want to derive the geometric albedo $p_V$, we use $V$ band photometry when available, otherwise we use $R$ band photometry and convert to $V$ band magnitudes using an averaged $\langle V-R\rangle=0.567\pm0.118$ color index\footnote[3]{The applied Plutino V-R color agrees within 1~$\sigma$ with our sample mean of 0.51$\pm$0.12 (cf. Table \ref{tab:optical_photometry}).}, and a correction for the intrinsic solar color index of $(V-R)_{\sun} = 0.36$ \citep{Doressoundiram2008}. $\langle V-R\rangle$ is based on observations of 41 Plutinos and is adopted from \citet[][data as of July 13, 2010]{HainautMBOSS}.

If available, we give priority to data drawn from refereed publications, due to the existence of photometric uncertainties and the better calibration. All available apparent magnitudes for a target are converted to $m_V(1,1)$ magnitudes and plotted as a function of the phase angle. For targets where the data show a clear trend of $m_V(1,1)$ vs. $\alpha$, we use a weighted linear regression analysis to fit a line to the data, yielding the linear phase coefficient. In the case of high scatter, a canonical $\beta = 0.10\pm0.04$ mag/$\degr$ is assumed, based on data given by \citet{Belskaya2008} ($V$ and $R$ band data, excluding Pluto). $H$ then represents the average of the values determined from each $m_V(1,1)$. MPC photometric data are usually highly scattered due to the coarse photometry, so the fixed-$\beta$ technique is always applied if only MPC data are available. The results of all computations are given in Table \ref{tab:optical_photometry}, which also gives lightcurve amplitude information and color indices, as far as available. Due to the usage of a two-visit observation strategy, the determined Herschel flux densities are a combination of flux densities taken at two different points in time. Hence, it is not trivial to correct both Herschel measurements for lightcurve effects. However, lightcurve effects are already included in the uncertainty of the absolute magnitude $H$, assuming that the optical measurements are randomly sampled with respect to the individual lightcurve. In cases where the uncertainty of $H$ is smaller than half the peak-to-peak lightcurve amplitude, we account for lightcurve effects by a quadratic addition of the latter value to the uncertainty of $H$. This was only necessary in three cases (cf. Table \ref{tab:optical_photometry}).



\section{Thermal Modeling}
\label{label:thermal_modeling}

By combining thermal-infrared and optical data, the physical properties of an object can be estimated using a thermal model. The disk-integrated thermal emission at wavelength $\lambda$ of a spherical model asteroid is calculated from the surface temperature distribution $T(\theta, \varphi)$ as
\begin{equation}
F(\lambda) = \epsilon d^2/\Delta^2 \int \int B(\lambda, T(\theta, \varphi)) \cos^2 \theta \, \cos(\varphi-\alpha) \,d\varphi\,d\theta,  \label{eqn:flux}
\end{equation}
where $\epsilon$ is the emissivity, $d$ the object's diameter, $\Delta$ its distance from the observer, $\alpha$ is the solar phase angle and $\theta$ and $\varphi$ are surface latitude and longitude measured from the subsolar point, respectively. $B(\lambda, T)$ is the Planck equation, $B(\lambda, T) = 2hc^2/\lambda^5 \: [\exp (hc/(\lambda k_B T))-1]^{-1}$, where $T$ is the temperature, $h$ Planck's constant, $c$ the velocity of light and $k_B$ Boltzmann's constant. The disk-integrated thermal emission is then fitted to the thermal-infrared data by variation of the model surface temperature distribution $T$. The body's temperature distribution depends on parameters such as albedo, thermal inertia, surface roughness, observation geometry, spin axis orientation and rotation period. In the modeling process, we take into account the individual observation geometries of each flux density measurement.

One of the first and most simple models is the Standard Thermal Model (STM, cf. \citet{Lebofsky1989} and references therein), which assumes the model body to be of spherical shape, non-rotating and/or having zero thermal inertia, with a smooth surface and observed at phase angle $\alpha = 0$. The surface temperature distribution is described by an instantaneous thermal equilibrium between emitted thermal radiation and absorbed sunlight. 

In this work we make use of the Near-Earth Asteroid Thermal Model (NEATM, \citet{Harris1998}), which is, despite its name, applicable to atmosphereless TNOs \citep[][the latter use a similar approach with their hybrid STM]{Mueller2010, Stansberry2008}. The main difference of the NEATM compared to the STM is the use of a variable beaming factor $\eta$, which adjusts the subsolar temperature 
\begin{equation}
 T_{SS} = [(1-A)S_{\sun}/(\epsilon \sigma \eta r^2)]^{1/4}, \label{eqn:T_SS}
\end{equation}
where $A$ is the Bond albedo, $S_{\sun}$ the solar constant, $\sigma$ the Stefan-Boltzmann constant and $r$ the object's heliocentric distance. $A$ is related to the geometric albedo, $p_V$, via $A \sim A_V = qp_V$ \citep{Lebofsky1989}. We use a phase integral $q = 0.336~p_V+0.479$ \citep{Brucker2009}. $\eta$ accounts for thermal inertia and surface roughness in a first-order approximation and can serve as a free parameter being derived as part of the fitting process, which usually provides better results than assuming a fixed value of $\eta$. Small values of $\eta$ ($\eta<1$) imply higher surface temperatures compared to that of a Lambertian surface, for instance due to surface roughness or porosity. High values of $\eta$ ($\eta>1$) lead to a reduction of the model surface temperature, mimicking the effect of thermal inertia. 
In contrast to the STM, the NEATM accounts for phase angles $\neq 0$ (as shown in Equation \ref{eqn:flux}). In the case of TNOs, however, this aspect is unimportant, since these objects are always observed at very low phase angles, due to their large heliocentric distances.  

Whenever Spitzer data are available we combine them with PACS data. The combination of the data has a large impact on the reliability and accuracy of the model output. The peak emission wavelength in flux density units (Jy) of typical Plutino orbits ranges from 80\,$\mu m$ to 100\,$\mu m$ (compare model SEDs in Figures \ref{fig:plots}, \ref{fig:plots2}, and \ref{fig:plots3}). Hence, PACS band wavelengths (70~$\mu m$, 100~$\mu m$ and 160~$\mu m$) are located in the flat peak plateau and the shallow Rayleigh-Jeans tail of the Planck function, respectively, and therefore constrain the model SED and $\eta$ insufficiently. Adding a Spitzer 24~$\mu m$ band measurement, which is located in the steeper Wien-slope, improves the ability to constrain the model SED and allows for the determination of reliable estimates of $\eta$. Therefore, we adopt floating-$\eta$ fits if additional Spitzer data are available and rely on floating-$\eta$ fits if only Herschel data are available. However, in some cases of poor data quality or whenever data from different parts of the lightcurve or from very different observation geometries are combined, the floating-$\eta$ method leads to $\eta$ values which are too high or too low and therefore unphysical (the range of physically meaningful $\eta$ values is $0.6 \leq \eta \leq 2.6$\footnote[4]{The physical range of $\eta$ values were probed by using NEATM to model the two extreme cases of (1) a fast rotating object of high thermal inertia with low surface roughness and (2) a slow rotating object of low thermal inertia with high surface roughness, which yield the upper and lower $\eta$ limit, respectively. Fluxes for both cases were determined using a full thermophysical model \citep{Lagerros1996, Lagerros1997, Lagerros1998, Mueller2007}, in which surface roughness is implemented via spherical cratering. The degree of surface roughness is determined via the opening angle (0\degr-180\degr) and surface density (0.0--1.0) of the craters.}). In such cases, we also apply the fixed-$\eta$ method, which makes use of a canonical value of $\eta = 1.20 \pm 0.35$ derived by \citet{Stansberry2008} from Spitzer observations of a sample of Centaurs and TNOs. The validity of this approach is discussed in Section \ref{ref:results}. We treat non-detections with a $1\,\sigma$ uncertainty as $0\pm\sigma$ detections in our models.

Throughout this work we assume the surface emissivity $\epsilon$ to be constant at a value of 0.9 for all wavelengths, which is based on laboratory measurements of silicate powder \citep{Hovis1966} and a commonly adopted approximation for small bodies in the Solar System. 

\begin{figure}[t]
 \centering
 \resizebox{\hsize}{!}{\includegraphics{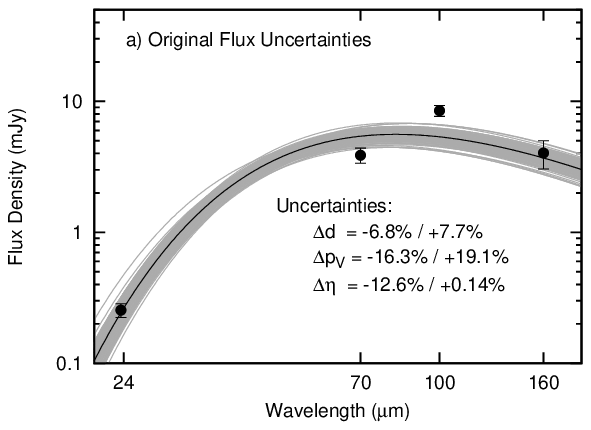}} \\ \resizebox{\hsize}{!}{\includegraphics{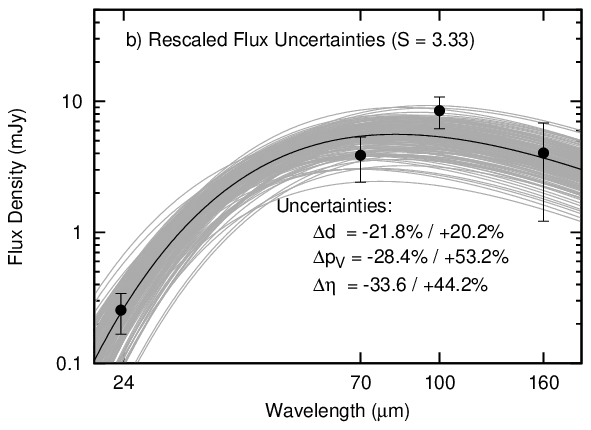}}
 \caption{Illustration of the rescaled flux density uncertainties technique using synthetic data. {\bf a)} Filled circles represent the flux densities at the different wavelengths with original flux density uncertainties. The dark line represents the best fit line ($\chi^2_{red}=11.09$); the grey lines are fit lines of 100 out of 1000 synthetical bodies generated for the MC method. The bad quality of the fit is obvious, since the 70 and 100~$\mu m$ datapoints are hardly fit by any of the single fit lines. Therefore, the reliability of the uncertainties of the output parameters is questionable. {\bf b) } By rescaling the error bars by a factor of $3.33=\sqrt{11.09}$, the broadness of the distribution of single fit lines is increased, which covers now all datapoints. As a result, the output parameter fractional uncertainties are increased and therefore more realistic.}
 \label{fig:chi2montecarlo} 
\end{figure}

subsection{Uncertainty Assessment}

In order to derive uncertainty estimates for diameter and albedo, we make use of a Monte Carlo (MC) simulation in which a sample of 1000 randomized synthetic bodies is created by variation of the observed flux density, the $\eta$ value, and the absolute magnitude $H$. The randomized parameters follow a normal distribution centered on the nominal value, within the limits of the respective uncertainties. The uncertainties in flux density and $H$ are taken from Table \ref{tab:PACS_observations} and Table \ref{tab:optical_photometry}, respectively. In the case of floating-$\eta$ fits, $\eta$ follows as a result of the modeling process, whereas in the case of fixed-$\eta$ fits, the value of $\eta\pm0.35$, determined by \citet{Stansberry2008} is applied. 

Uncertainties of a specific parameter are derived from the ensemble of modeled synthetic bodies as the upper and lower values that include $68.2\%$ of the ensemble (centered on the median), respectively. The uncertainties are usually aligned asymmetrically around the median, since most diameter and albedo distributions do not strictly follow a normal distribution. This method was introduced by \citet{Mueller2011}. Finally, we combine the uncertainties determined through the MC method with the respective parameter values provided by the best model fit. 

In order to improve uncertainty estimates, we make use of the rescaled uncertainties approach, introduced by \citet{SantosSanz2011}. The motivation for this technique is the fact, that uncertainty estimates resulting from the MC method might be unrealistically small in cases in which the model fits the observational data only poorly. As a measure for the fit quality we use the reduced $\chi^2$, which is a result of the fitting process. It is defined as 
\begin{equation}
 \chi^2_{red} = \frac{1}{\mbox{n.d.f.}} \sum_{i} \left [ \frac{F_{obs,i}-F_{model,i}}{\sigma_{obs,i}} \right ]^2, \label{eqn:chi2}
\end{equation} 
with $F_{obs,i}$ and $F_{model,i}$ being the observed and modeled flux densities of the different observations $i$, respectively, and $\sigma_{obs,i}$ being the respective observational flux density uncertainty. $\chi^2_{red}$ is normalized by the number of degrees of freedom (n.d.f.), which equals the number of datapoints $-2$ in the case of a floating-$\eta$ fit and the number of datapoints $-1$ in the case of a fixed-$\eta$ method, respectively. A poor fit results in $\chi^2_{red} \gg 1$. 

The idea is to increase the flux density uncertainties uniformly by a factor $S$, which leads to an increase in variation of the flux densities of the synthetic bodies. The value of $S$ is determined in such a way as to provide by definition the best possible model fit, i.e. $\chi^2_{red} = 1$, which is achieved by choosing $S = \sqrt{\chi^2_{red,min}}$, with $\chi^2_{red,min}$ being the minimum $\chi^2_{red}$ resulting from fitting the unaltered observational data. Hence, the rescaled flux density uncertainties are $\hat\sigma_{obs} = S \, \sigma_{obs}$ and the resulting $\hat\chi^2_{red,min}$, which is obtained by replacing $\sigma_{obs}$ with $\hat\sigma_{obs}$ in equation \ref{eqn:chi2}, is by definition unity. Diameter uncertainties are dominated by flux uncertainties and the quality of the fit, whereas albedo uncertainties are dominated by the uncertainty of $H$. The rescaled uncertainties are used only in the MC method to determine the uncertainties of the physical parameters. In order to obtain the best fit, the original uncertainties are applied. 

The method is illustrated in Figure \ref{fig:chi2montecarlo} using synthetic data. Figure \ref{fig:chi2montecarlo}~a) shows the original flux densities with error bars of the size of the original uncertainties. The dark line represents the best NEATM fit (floating-$\eta$ method) of the dataset, whereas each grey line symbolizes the best fit SED of one of 100 of the 1000 randomized synthetic bodies. It is obvious that the distribution of single fit curves misses the 100~$\mu m$ PACS band completely and hardly fits the 70~$\mu m$ datapoint. Hence, the fit quality is bad, which results in an underestimation of the model output uncertainties. 
By rescaling the uncertainties of the flux densities, the broadness of the distribution of single fit lines increases, as depicted in Figure \ref{fig:chi2montecarlo}~b). As a consequence of the rescaling, all datapoints are covered by the set of lines and the derived uncertainties are significantly increased, leading to much more realistic uncertainty estimates.


\begin{table}[!t]
 \caption{Modeling results sorted by modeling technique. `Data' denotes the dataset on which the modeling results are based: `H' equals Herschel data, `HS' equals Herschel and Spitzer data, respectively; diameter, geometric albedo and $\eta$ are given including respective upper and lower uncertainties. We show all succesful model fit results in order to allow for a comparison of the reliability of the techniques and the data sets.}
 \centering
 \begin{tabular}{rcccc}
 \hline \hline  \\[-7pt]
 Object & Data & $d$ [km] & $p_V$ & $\eta$\\
 \hline \\ [-7pt]
     \multicolumn{5}{l}{{\bf floating-\boldmath$\eta$ fits:}} \\ [+2.5pt]
   1999~TC36        & HS & $   393.1_{   -26.8}^{+    25.2}$ & $ 0.079_{-0.011}^{+ 0.013}$ & $ 1.10_{-0.08}^{+ 0.07}$  \\ [+2.5pt]
  2000~GN171        & HS & $   147.1_{   -17.8}^{+    20.7}$ & $ 0.215_{-0.070}^{+ 0.093}$ & $ 1.11_{-0.21}^{+ 0.24}$  \\ [+2.5pt]
   2002~VE95        & HS & $   249.8_{   -13.1}^{+    13.5}$ & $ 0.149_{-0.016}^{+ 0.019}$ & $ 1.40_{-0.11}^{+ 0.12}$  \\ [+2.5pt]
   2002~XV93        & H  & $   451.3_{   -64.4}^{+    61.7}$ & $ 0.060_{-0.025}^{+ 0.042}$ & $ 0.72_{-0.24}^{+ 0.30}$  \\ [+2.5pt]
   2002~XV93        & HS & $   549.2_{   -23.0}^{+    21.7}$ & $ 0.040_{-0.015}^{+ 0.020}$ & $ 1.24_{-0.06}^{+ 0.06}$  \\ [+2.5pt]
   2003~AZ84$^a$    & HS & $   727.0_{   -66.5}^{+    61.9}$ & $ 0.107_{-0.016}^{+ 0.023}$ & $ 1.05_{-0.15}^{+ 0.19}$  \\ [+2.5pt]
    2003~VS2$^a$    & HS & $   523.0_{   -34.4}^{+    35.1}$ & $ 0.147_{-0.043}^{+ 0.063}$ & $ 1.57_{-0.23}^{+ 0.30}$  \\ [+2.5pt]
   2004~EW95        & H  & $   265.9_{   -45.2}^{+    48.3}$ & $ 0.053_{-0.022}^{+ 0.035}$ & $ 0.93_{-0.42}^{+ 0.55}$  \\ [+2.5pt]
  2004~PF115        & H  & $   406.3_{   -85.3}^{+    97.6}$ & $ 0.113_{-0.042}^{+ 0.082}$ & $ 0.84_{-0.40}^{+ 0.61}$  \\ [+2.5pt]
   2004~UX10        & H  & $   361.0_{   -93.5}^{+   124.2}$ & $ 0.172_{-0.078}^{+ 0.141}$ & $ 0.96_{-0.56}^{+ 0.98}$  \\ [+2.5pt]
  2006~HJ123        & H  & $   283.1_{  -110.8}^{+   142.3}$ & $ 0.136_{-0.089}^{+ 0.308}$ & $ 2.48_{-1.91}^{+ 3.92}$  \\ [+2.5pt]
        Huya        & H  & $   395.7_{   -34.7}^{+    35.7}$ & $ 0.100_{-0.018}^{+ 0.022}$ & $ 0.75_{-0.19}^{+ 0.22}$  \\ [+2.5pt]
        Huya        & HS & $   438.7_{   -25.2}^{+    26.5}$ & $ 0.081_{-0.011}^{+ 0.011}$ & $ 0.89_{-0.06}^{+ 0.06}$  \\ [+2.5pt]
     \hline \\ [-7pt]
     \multicolumn{5}{l}{{\bf fixed-\boldmath$\eta$ fits:}} \\ [+2.5pt]
   1996~TP66        & HS & $   154.0_{   -33.7}^{+    28.8}$ & $ 0.074_{-0.031}^{+ 0.063}$  \\ [+2.5pt]
   1999~TC36        & H  & $   407.5_{   -38.7}^{+    38.9}$ & $ 0.073_{-0.014}^{+ 0.019}$  \\ [+2.5pt]
   1999~TC36        & HS & $   428.3_{  -112.2}^{+    81.7}$ & $ 0.066_{-0.023}^{+ 0.064}$  \\ [+2.5pt]
  2000~GN171        & H  & $   150.0_{   -19.1}^{+    19.7}$ & $ 0.207_{-0.070}^{+ 0.105}$  \\ [+2.5pt]
  2000~GN171        & HS & $   155.1_{   -27.3}^{+    20.4}$ & $ 0.193_{-0.075}^{+ 0.141}$  \\ [+2.5pt]
   2001~KD77        & H  & $   232.3_{   -39.4}^{+    40.5}$ & $ 0.089_{-0.027}^{+ 0.044}$  \\ [+2.5pt]
  2001~QF298        & H  & $   408.2_{   -44.9}^{+    40.2}$ & $ 0.071_{-0.014}^{+ 0.020}$  \\ [+2.5pt]
   2002~VE95        & H  & $   237.0_{   -21.4}^{+    18.2}$ & $ 0.165_{-0.027}^{+ 0.037}$  \\ [+2.5pt]
   2002~VE95        & HS & $   226.3_{   -42.5}^{+    27.6}$ & $ 0.181_{-0.042}^{+ 0.113}$  \\ [+2.5pt]
  2002~VR128        & H  & $   448.5_{   -43.2}^{+    42.1}$ & $ 0.052_{-0.018}^{+ 0.027}$  \\ [+2.5pt]
  2002~VU130        & H  & $   252.9_{   -31.3}^{+    33.6}$ & $ 0.179_{-0.103}^{+ 0.202}$  \\ [+2.5pt]
   2002~XV93        & H  & $   544.1_{   -55.6}^{+    47.8}$ & $ 0.041_{-0.017}^{+ 0.026}$  \\ [+2.5pt]
   2002~XV93        & HS & $   534.3_{  -124.2}^{+    69.6}$ & $ 0.042_{-0.021}^{+ 0.048}$  \\ [+2.5pt]
   2003~AZ84$^a$    & H  & $   771.0_{   -77.9}^{+    78.9}$ & $ 0.095_{-0.019}^{+ 0.026}$  \\ [+2.5pt]
   2003~AZ84$^a$    & HS & $   766.8_{   -97.8}^{+    58.9}$ & $ 0.096_{-0.016}^{+ 0.035}$  \\ [+2.5pt]
  2003~UT292        & H  & $   186.6_{   -20.3}^{+    17.3}$ & $ 0.092_{-0.049}^{+ 0.091}$  \\ [+2.5pt]
    2003~VS2        & H  & $   483.1_{   -50.0}^{+    37.0}$ & $ 0.172_{-0.056}^{+ 0.092}$  \\ [+2.5pt]
   2003~VS2$^a$     & H  & $   465.7_{   -51.0}^{+    45.7}$ & $ 0.185_{-0.062}^{+ 0.091}$  \\ [+2.5pt]
    2003~VS2$^a$    & HS & $   466.9_{   -49.9}^{+    38.6}$ & $ 0.184_{-0.065}^{+ 0.110}$  \\ [+2.5pt]
   2004~EW95        & H  & $   291.1_{   -25.9}^{+    20.3}$ & $ 0.044_{-0.015}^{+ 0.021}$  \\ [+2.5pt]
  2004~PF115        & H  & $   468.2_{   -49.1}^{+    38.6}$ & $ 0.123_{-0.033}^{+ 0.043}$  \\ [+2.5pt]
   2004~UX10        & H  & $   398.1_{   -39.3}^{+    32.6}$ & $ 0.141_{-0.031}^{+ 0.044}$  \\ [+2.5pt]
  2006~HJ123        & H  & $   216.4_{   -34.2}^{+    29.7}$ & $ 0.281_{-0.152}^{+ 0.259}$  \\ [+2.5pt]
        Huya        & H  & $   461.1_{   -41.4}^{+    31.2}$ & $ 0.073_{-0.011}^{+ 0.017}$  \\ [+2.5pt]
        Huya        & HS & $   561.5_{  -111.4}^{+    82.8}$ & $ 0.049_{-0.017}^{+ 0.041}$  \\ [+2.5pt]
      Pluto/Charon  & H  & $  2119.9_{  -182.2}^{+   164.5}$ & $ 0.730_{-0.153}^{+ 0.162}$  \\ [+2.5pt]
 \hline
 \end{tabular}
 \tablefoot{In the case of all fixed-$\eta$ modeling approaches, $\eta=1.20\pm0.35$ was applied.
   \tablefoottext{a}{Using averaged lightcurve observations.}
 }
 \label{tab:all_results}
\end{table}

\section{Results and Discussion}

An overview of all modeling results is given in Table \ref{tab:all_results}. Where possible, floating and fixed-$\eta$ modeling approaches were applied based on Herschel-only and combined Herschel and Spitzer datasets, in order to test the consistency of the results. The objects' SEDs determined by the different models are plotted in Figures \ref{fig:plots}, \ref{fig:plots2} and \ref{fig:plots3}.

In the following sections we test the validity of the $\eta=1.2$ assumption for the Plutino sample, and discuss the modeling results of the individual objects. After that, the results of the sample as a whole are discussed and used to determine a cumulative size distribution of the Plutino population. Finally, correlations between the different physical, photometric and dynamical parameters and the presence of ices are examined.

\subsection{Test of the Fixed-$\eta$ Assumption}
\label{ref:results}

\begin{figure*}[!t]
 \centering
 \begin{tabular}{cc}
    \includegraphics[width=7.0cm]{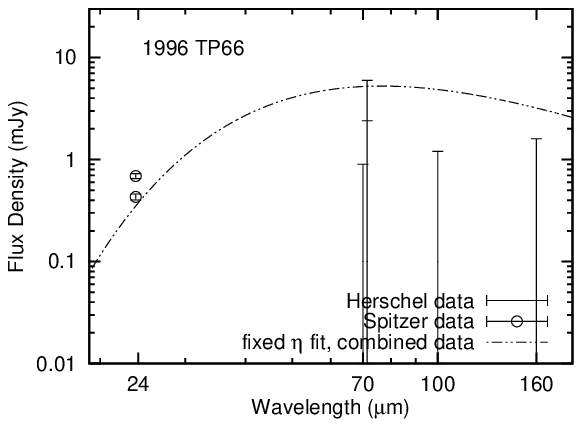}  & \includegraphics[width=7.0cm]{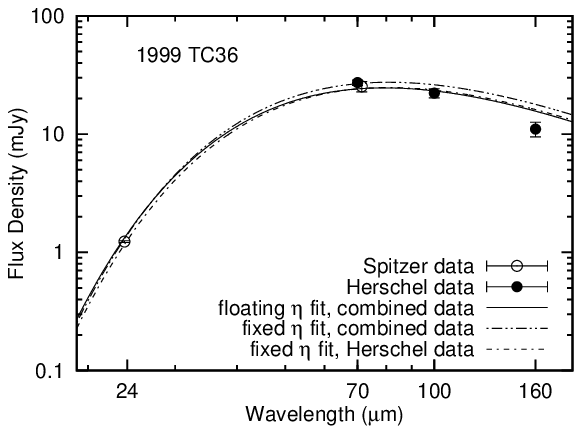} \\
    \includegraphics[width=7.0cm]{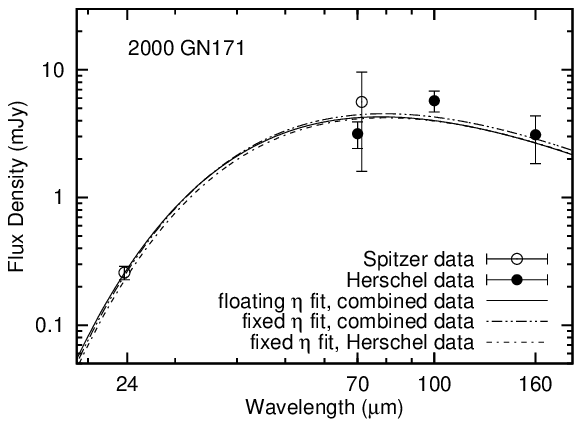} & \includegraphics[width=7.0cm]{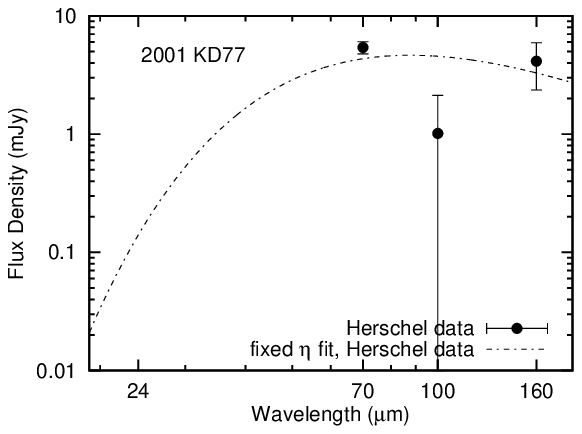} \\
    \includegraphics[width=7.0cm]{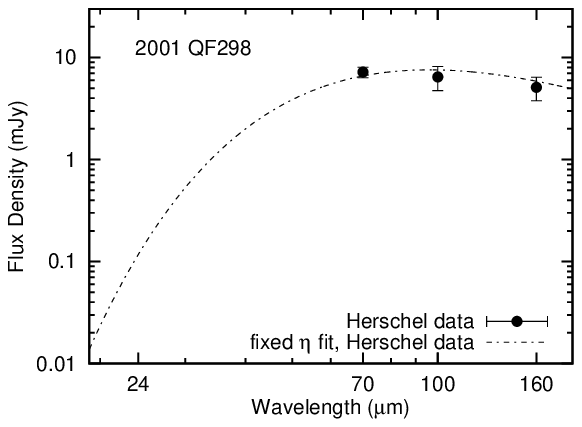} & \includegraphics[width=7.0cm]{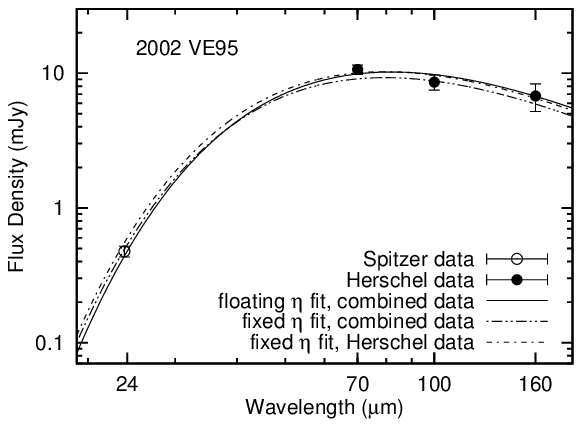} \\
    \includegraphics[width=7.0cm]{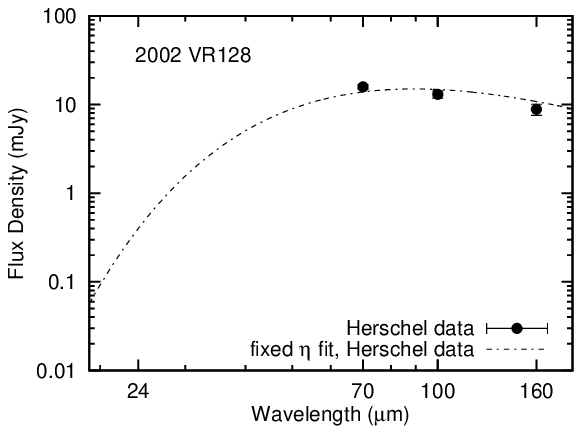} & \includegraphics[width=7.0cm]{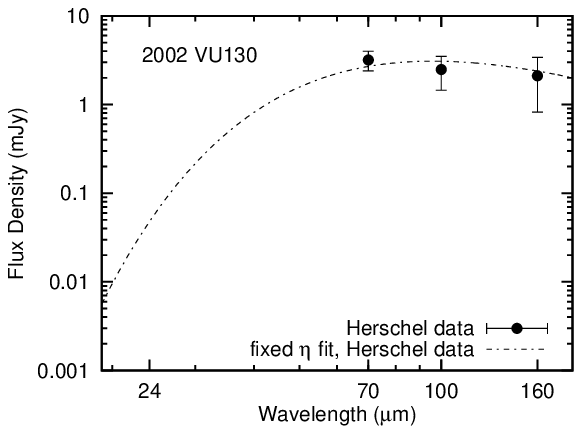} \\
 \end{tabular} 
 \caption{Model fits of the Plutino sample using different modeling approaches and different data samples: the solid and dashed lines are floating-$\eta$ fits based on combined Herschel and Spitzer and Herschel-only data, respectively; the dash-dotted and dash-dot-dotted lines represent fixed-$\eta$ fits of combined Herschel and Spitzer and Herschel-only data samples, respectively; datapoints at 23.68~$\mu m$ and 71.42~$\mu m$ are Spitzer MIPS data (open circles), those at 70~$\mu m$, 100~$\mu m$ and 160~$\mu m$ are Herschel PACS data (filled circles). The model is fit to each flux measurement using the circumstances appropriate for the epoch of that observation. In order to simplify the figures, we have normalized the Spitzer flux densities to the epoch of the Herschel observations using the ratio of the best-fit model flux densities for the Herschel and Spitzer epoch, and then plotted the measured value times this ratio. Uncertainties are rescaled similarly, preserving the SNR for the Spitzer data.}
 \label{fig:plots} 
\end{figure*}

\begin{figure*}[t]
 \centering
 \begin{tabular}{cc}
    \includegraphics[width=7.0cm]{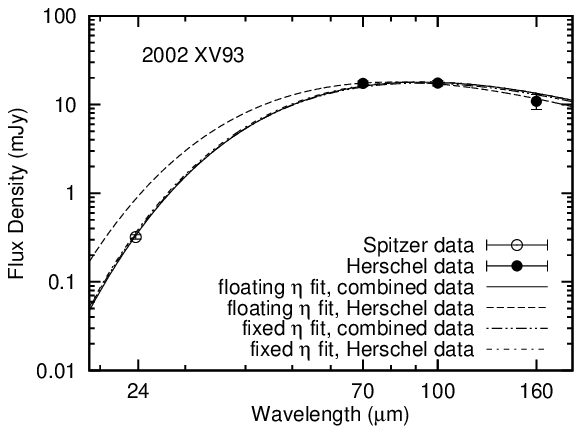}  & \includegraphics[width=7.0cm]{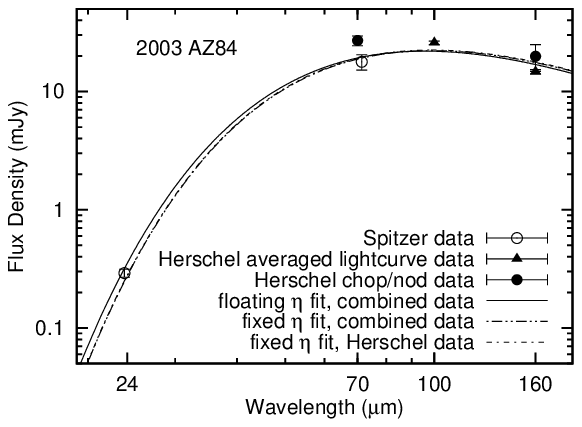} \\
    \includegraphics[width=7.0cm]{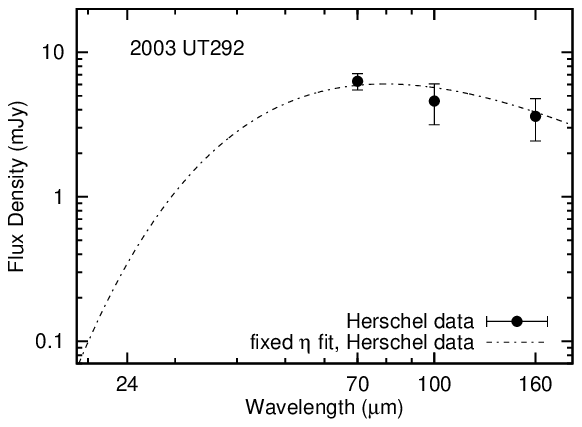} & \includegraphics[width=7.0cm]{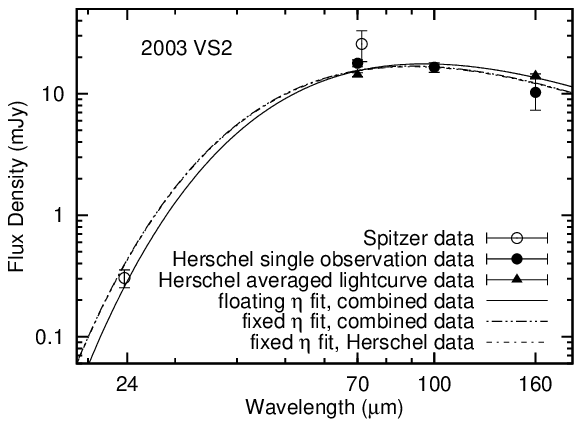}  \\
    \includegraphics[width=7.0cm]{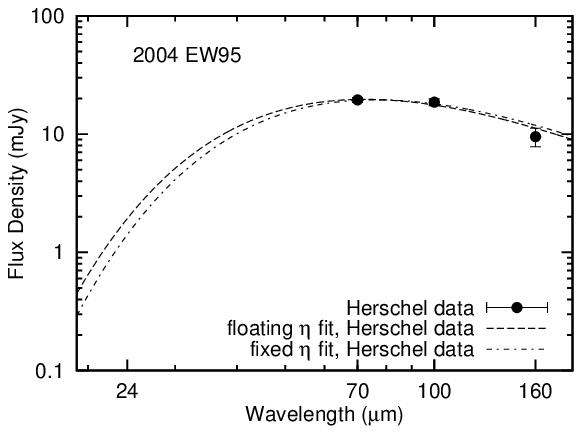}  & \includegraphics[width=7.0cm]{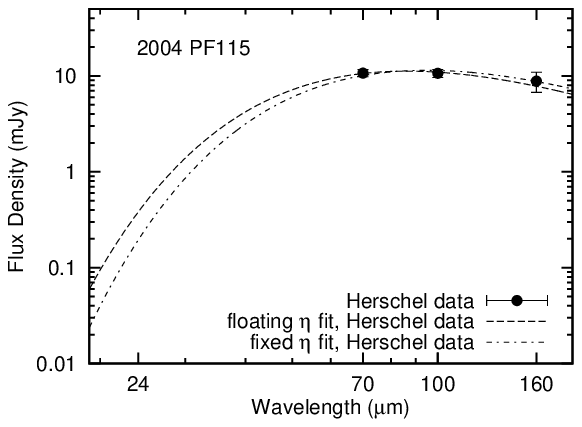} \\
    \includegraphics[width=7.0cm]{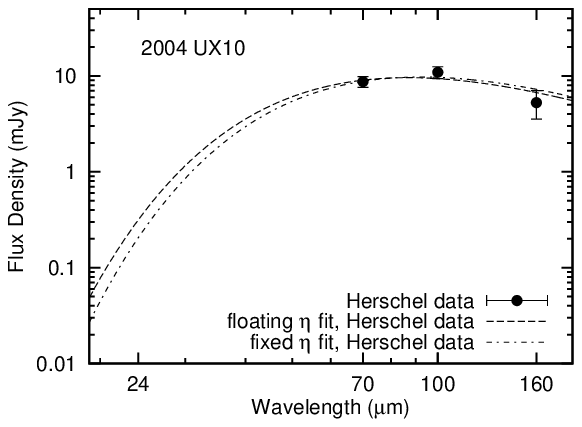}  & \includegraphics[width=7.0cm]{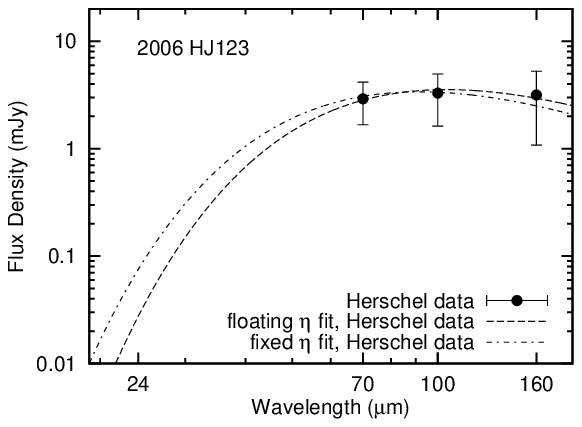} \\
 \end{tabular} 
 \caption{Continuation of Figure \ref{fig:plots}.}
 \label{fig:plots2} 
\end{figure*}

\begin{figure*}[t]
 \centering
 \begin{tabular}{cc}
    \includegraphics[width=7.0cm]{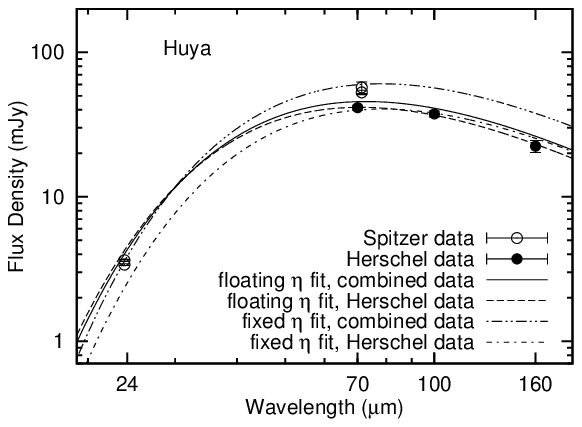}      & \includegraphics[width=7.0cm]{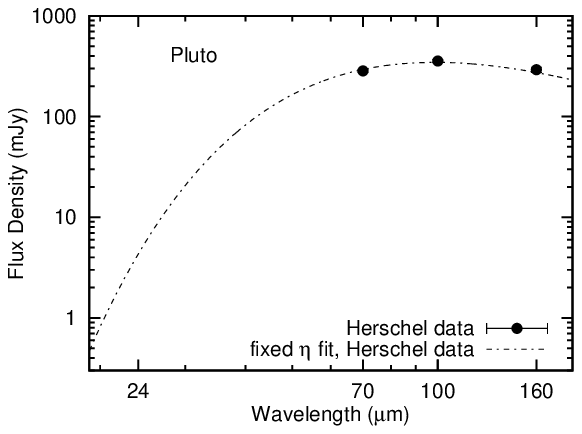} \\
 \end{tabular} 
 \caption{Continuation of Figure \ref{fig:plots2}.}
 \label{fig:plots3} 
\end{figure*}

We were able to derive floating-$\eta$ fits for 12 targets; we adopted the 7 which are based on combined Herschel and Spitzer data. The top section of Table \ref{tab:all_results} gives the results of all converged floating-$\eta$ fits. The weighted mean\footnote[5]{The weighted mean and its uncertainties were determined using the equation given in Table \ref{tab:optical_photometry}.} $\eta$ determined from the 7 adopted floating-$\eta$ model fits yields $\langle \eta \rangle = 1.11_{-0.19}^{+0.18}$, which is consistent with the adopted value of $\eta=1.20\pm0.35$ \citep{Stansberry2008} applied in the fixed-$\eta$ approach. The root mean square fractional residuals in the results of the fixed-$\eta$ method between using $\eta=1.20$ and $\langle \eta \rangle = 1.11$ is 5~\% and 10~\% for diameter and albedo, respectively. However, the average fractional uncertainties in diameter and albedo of all adopted results are $\pm$~10~\% and +49/-31~\%, respectively, which is larger than the residuals. Therefore, we continue to adopt $\eta=1.20$ for the fixed-$\eta$ approach.

\subsection{Discussion of Individual Targets}
\label{ref:discussion}

\noindent{\it 1996~TP66}: Despite the availability of both Herschel and Spitzer data, the floating-$\eta$ approach was not applicable for 1996~TP66, because of the non-detections in the PACS 70, 100 and 160~$\mu m$ bands. The best fit fixed-$\eta$ solution is largely constrained by the two 24~$\mu m$ MIPS flux densities. Therefore, our results ($d=154.0_{-33.7}^{+28.8}$~km , $p_V=0.074_{-0.031}^{+ 0.063}$, based on $\eta=1.20$) barely differ from those of \citet{Stansberry2008}, who find $d=160^{+45}_{-45}$~km and $p_V=0.074^{+0.07}_{-0.03}$ using $\eta=1.20^{+0.35}_{-0.35}$, solely based on the 24~$\mu m$ MIPS flux densities. The reason for the non-detections is unclear. However, the fact that no clear detection was possible in the wavelength regime of 70~$\mu m$ and above using two different instruments (PACS and MIPS) and at different epochs points to a real effect. We based the observation planning on $p_V=0.07$, which agrees well with the best fit model.

\noindent{\it 1999~TC36}: This target was well detected in all three PACS bands. In combination with MIPS flux densities, a good floating-$\eta$ fit was possible here, which misses the PACS 160~$\mu m$ band flux density. This might hint to a more complex surface albedo distribution or a wavelength-dependency of the emissivity. Our adopted diameter and albedo estimates, $393.1_{-26.8}^{+25.2}$~km and $0.079_{-0.011}^{+0.013}$, respectively, agree with earlier estimates by \citet{Stansberry2008}: $d=414.6_{-38.2}^{+38.8}$~km and $p_V=0.072_{-0.012}^{+0.015}$. 1999~TC36 is a triple component system \citep{Trujillo2002, Jacobson2007, Benecchi2010}, consisting of two similarly sized central components $A_1$ and $A_2$ and a more distant secondary component, $B$. The nature of this system makes it possible to determine the mean density of the system and, using optical flux differences between the single components, the sizes of the components. We revisit the calculations performed by \citet{Stansberry2006} and \citet{Benecchi2010}. Our newly derived estimates of the sizes, based on the flux differences found by \citet{Benecchi2010}, are: $d_{A1}=272_{-19}^{+17}$~km, $d_{A2}=251_{-17}^{+16}$~km and $d_B=132_{-9}^{+8}$~km. Using these diameters, we are able to improve the mean system density $\varrho=0.64_{-0.11}^{+0.15}\,\mbox{g cm}^{-3}$, which is somewhat higher than the earlier estimate $\varrho=0.54_{-0.21}^{+0.32}\,\mbox{g cm}^{-3}$ \citep{Benecchi2010}. Assuming material densities of $1.0 \leq \varrho_0 \leq 2.0\,\mbox{g cm}^{-3}$, we find porosities or fractional void space of 36--68\%. Our calculations are based on the assumptions of spherical shape of each component and the same albedo for each component.

\noindent{\it 2000~GN171}: Both fixed and floating-$\eta$ model fits based on the combined Herschel and Spitzer datasets were possible, leading to good fits with similar outcomes. We adopt the floating-$\eta$ result $d=147.1_{-17.8}^{+20.7}$~km, $p_V=0.215_{-0.070}^{+0.093}$, and $\eta=1.11_{-0.21}^{+0.24}$, which differs significantly from an earlier estimate by \citet{Stansberry2008}, who determined $d=321^{+57.4}_{-54.2}$\,km, $p_V=0.057^{+0.025}_{-0.016}$ and $\eta=2.32^{+0.46}_{-0.43}$. The large discrepancy stems from an overestimated MIPS 70\,$\mu m$ band flux density in the \citet{Stansberry2008} data, which has been revised for this work. By adding the Herschel PACS data, the SED is better constrained, leading to a more reliable result. 2000~GN171 is either a Roche binary or a Jacobi ellipsoid \citep{Lacerda2007}. With a lightcurve amplitude of 0.61 mag \citep{Sheppard2002}, the size of the companion or the ellipticity of the body is significant. Hence, the real diameter of 2000~GN171, the smallest Plutino in our sample, might be even smaller.

\noindent{\it 2001~KD77}: We modeled this target solely on the basis of PACS data, since existing MIPS flux density measurements suffer from a large positional uncertainty. For this faint target, the fixed-$\eta$ fit mainly relies on the 70~$\mu m$ band measurement, which has the highest SNR of all bands. Figure \ref{fig:plots} clearly shows the mismatch of the 100~$\mu m$ band measurement, which is scarcely detectable on the sky subtracted maps. Due to its large uncertainty, this datapoint is nearly neglected by the fixed-$\eta$ fit, and therefore scarcely impacts the modeling results. The reason for the low 100~$\mu m$ flux density is unknown.

\noindent{\it 2002~VE95}, {\it 2002~XV93}, {\it 2003~AZ84}, {\it 2003~VS2} and {\it Huya}: These objects show a good to excellent agreement of both the different model technique fits compared to each other, as well as the model SED compared with the different measured flux densities. For all five targets combined Herschel and Spitzer flux densities are available and we adopt the floating-$\eta$ modeling results. 2003~AZ84 and 2003~VS2 data include averaged lightcurve data, which strongly constrain the respective SEDs, due to their small uncertainties. Earlier estimates of 2003~VS2 by \citet{Stansberry2008} based on Spitzer data suggested $d=725.2_{-187.6}^{+199.0}$~km and $p_V=0.058_{-0.022}^{+0.048}$ using $\eta=2.00_{-0.51}^{+0.54}$, which is significantly different from our results $d=523.0_{-34.4}^{+35.1}$~km, $p_V=0.147_{-0.043}^{+0.063}$, based on $\eta=1.57_{-0.23}^{+0.30}$. The discrepancy might stem from the high 70~$\mu m$ MIPS flux density value, which, together with the 24~$\mu m$ MIPS data point, suggests a steeper Wien slope of the spectral energy distribution compared to the combined Herschel and Spitzer data set. 2003~AZ84 was observed before by Herschel during the Science Demonstration Phase using the chop/nod technique \citep{Mueller2010}, yielding $d=896\pm55$ and $p_V=0.065\pm0.008$, using $\eta=1.31\pm0.08$, which differs from our results $d=727.0_{-66.5}^{+61.9}$~km, $p_V=0.107_{-0.016}^{+0.023}$, based on $\eta=1.05_{-0.15}^{+0.19}$. Despite the smaller diameter determined using the averaged lightcurve data, 2003~AZ84 is still the largest Plutino in our sample, apart from Pluto. 2003~AZ84 has a moon, which is $5.0\pm0.3$ mag fainter than the primary \citep{Brown2007}. This large difference in magnitude suggests that the thermal flux of the companion is negligible. The modeling results of Huya vary significantly depending on the model technique and the dataset. A visual inspection of the model fits shows that the floating-$\eta$ approach based on the combined data set matches the measured SED best and results in $d=438.7_{-25.2}^{+26.5}$ and $p_V=0.081_{-0.011}^{+ 0.011}$, based on $\eta=0.89_{-0.06}^{+0.06}$. Earlier estimates by \citet{Stansberry2008} based on Spitzer data suggest $d=532.6_{-24.4}^{+25.1}$ and $p_V=0.050_{-0.004}^{+0.005}$, based on $\eta=1.09_{-0.06}^{+0.07}$. 2002~XV93 has the lowest albedo in the sample.

\noindent{\it 2001~QF298}, {\it 2002~VR128}, {\it 2002~VU130}, {\it 2003~UT292}, {\it 2004~EW95}, {\it 2004~PF115}, {\it2004~UX10} and {\it 2006~HJ123}: For these targets we have to rely on Herschel-only data and therefore adopt the results of the fixed-$\eta$ technique. We were able to succesfully apply the floating-$\eta$ method to 2004~EW95, 2004~PF115, 2004~UX10 and 2006~HJ123 as well. Modeling results in Table \ref{tab:all_results} and the plots in Figures \ref{fig:plots} and \ref{fig:plots2} of the SEDs show a good agreement between the results of the different techniques, supporting the validity of the fixed-$\eta$ approach and the reliability of the Herschel flux measurements. Differences in the fit quality are induced by the different SNR values of the respective measurements. Both 2002~VU130 and 2006~HJ123 suffer from large albedo uncertainties, which is a result of the large uncertainty of their absolute magnitude $H$. 2006~HJ123 shows the highest albedo of our sample, excluding Pluto.

\begin{table}[b]
 \caption{Adopted modeling results. Columns as in Table \ref{tab:all_results}. $\eta$ values of floating-$\eta$ modeling results are in bold.}
 \centering
 \begin{tabular}{rcccc}
 \hline \hline \\ [-7pt]
 Object & Data & $d$ [km] & $p_V$ & $\eta$\\
 \hline \\ [-7pt]
   1996~TP66        & HS & $   154.0_{   -33.7}^{+    28.8}$ & $ 0.074_{-0.031}^{+ 0.063}$ &          $ 1.20_{-0.35}^{+ 0.35}$  \\ [+3pt]
   1999~TC36        & HS & $   393.1_{   -26.8}^{+    25.2}$ & $ 0.079_{-0.011}^{+ 0.013}$ & \boldmath$ 1.10_{-0.08}^{+ 0.07}$  \\ [+3pt]
  2000~GN171        & HS & $   147.1_{   -17.8}^{+    20.7}$ & $ 0.215_{-0.070}^{+ 0.093}$ & \boldmath$ 1.11_{-0.21}^{+ 0.24}$  \\ [+3pt]
   2001~KD77        & H  & $   232.3_{   -39.4}^{+    40.5}$ & $ 0.089_{-0.027}^{+ 0.044}$ &          $ 1.20_{-0.35}^{+ 0.35}$  \\ [+3pt]
  2001~QF298        & H  & $   408.2_{   -44.9}^{+    40.2}$ & $ 0.071_{-0.014}^{+ 0.020}$ &          $ 1.20_{-0.35}^{+ 0.35}$  \\ [+3pt]
   2002~VE95        & HS & $   249.8_{   -13.1}^{+    13.5}$ & $ 0.149_{-0.016}^{+ 0.019}$ & \boldmath$ 1.40_{-0.11}^{+ 0.12}$  \\ [+3pt]
  2002~VR128        & H  & $   448.5_{   -43.2}^{+    42.1}$ & $ 0.052_{-0.018}^{+ 0.027}$ &          $ 1.20_{-0.35}^{+ 0.35}$  \\ [+3pt]
  2002~VU130        & H  & $   252.9_{   -31.3}^{+    33.6}$ & $ 0.179_{-0.103}^{+ 0.202}$ &          $ 1.20_{-0.35}^{+ 0.35}$  \\ [+3pt]
   2002~XV93        & HS & $   549.2_{   -23.0}^{+    21.7}$ & $ 0.040_{-0.015}^{+ 0.020}$ & \boldmath$ 1.24_{-0.06}^{+ 0.06}$  \\ [+3pt]
   2003~AZ84$^a$    & HS & $   727.0_{   -66.5}^{+    61.9}$ & $ 0.107_{-0.016}^{+ 0.023}$ & \boldmath$ 1.05_{-0.15}^{+ 0.19}$  \\ [+3pt]
  2003~UT292        & H  & $   185.6_{   -18.0}^{+    17.9}$ & $ 0.067_{-0.034}^{+ 0.068}$ &          $ 1.20_{-0.35}^{+ 0.35}$  \\ [+3pt]
    2003~VS2$^a$    & HS & $   523.0_{   -34.4}^{+    35.1}$ & $ 0.147_{-0.043}^{+ 0.063}$ & \boldmath$ 1.57_{-0.23}^{+ 0.30}$  \\ [+3pt]
   2004~EW95        & H  & $   291.1_{   -25.9}^{+    20.3}$ & $ 0.044_{-0.015}^{+ 0.021}$ &          $ 1.20_{-0.35}^{+ 0.35}$  \\ [+3pt]
  2004~PF115        & H  & $   468.2_{   -49.1}^{+    38.6}$ & $ 0.123_{-0.033}^{+ 0.043}$ &          $ 1.20_{-0.35}^{+ 0.35}$  \\ [+3pt]
   2004~UX10        & H  & $   398.1_{   -39.3}^{+    32.6}$ & $ 0.141_{-0.031}^{+ 0.044}$ &          $ 1.20_{-0.35}^{+ 0.35}$  \\ [+3pt]
  2006~HJ123        & H  & $   216.4_{   -34.2}^{+    29.7}$ & $ 0.281_{-0.152}^{+ 0.259}$ &          $ 1.20_{-0.35}^{+ 0.35}$  \\ [+3pt]
        Huya        & HS & $   438.7_{   -25.2}^{+    26.5}$ & $ 0.081_{-0.011}^{+ 0.011}$ & \boldmath$ 0.89_{-0.06}^{+ 0.06}$  \\ [+3pt]
 \hline
 \end{tabular}
 \tablefoot{No adopted values are given for Pluto/Charon, since our model approaches are not applicable to these objects (cf. Section \ref{ref:discussion}).
   \tablefoottext{a}{using averaged lightcurve observations.}
 }
 \label{tab:adopted_results}
\end{table}

\noindent{\it Pluto}: Paradoxically, Pluto is a clear outlier from the Plutino population, being by far the largest and having the brightest surface. Besides that, Pluto was the first TNO discovered to have an atmosphere \citep{Hubbard1988, Brosch1995}, has a pronounced optical lightcurve caused by albedo variations, and has a moon half its own size. An additional complication is the effect of N$_2$ ice on Pluto's surface: sublimation and deposition of N$_2$ is accompanied by latent heat transport that results in much lower (higher) dayside (nightside) temperatures than would be expected on an airless body \citep[e.g.][]{Spencer1997}. Despite these facts, we attempt to treat Pluto like any other
Plutino and compare our modeling results to the known properties of Pluto and Charon, and discuss the reasons for discrepancies that emerge. 

The Pluto system has been subject to detailed study via, for instance, direct imaging, stellar occultation observations and thermal radiometry in the past. Mean hemispheric albedo ranges are $0.49$ to $0.66$ for Pluto and $0.36$ to $0.39$ for Charon \citep{Buie1997}.  Charon's diameter is well determined to be 1208~km \citep{Gulbis2006,Sicardy2006}. Pluto's diameter estimates suffer from uncertainties related to the presence of the atmosphere and range from about 2290--2400~km \citep{Tholen1997}. For the purpose of comparing with the results of our thermal models, we adopt a diameter of 2322~km \citep{Young2007}. These numbers lead to an equivalent diameter and albedo of the combined system of 2617~km and $0.44$--$0.73$, respectively.

We deliberately refrain from combining our Herschel data set with existing Spitzer measurements (for instance \citet{Lellouch2011}), due to the so far unexplained secular fading of the Pluto/Charon system \citep{Stansberry2009}. Instead, we rely on our original Herschel measurements only, and so we adopt the fixed-$\eta$ results: $d=2119.9_{-182.2}^{+164.5}$~km and $p_V=0.730_{-0.153}^{+0.162}$ (cf. Figure \ref{fig:plots3}). Our thermal estimate of the effective system diameter deviates by 3~$\sigma$ from the known effective system diameter given above; the albedo derived from our thermal modeling is only barely consistent with the maximum allowable albedo given above. Our attempt to apply a simple thermal model to Pluto shows the susceptibility of such models to violations of the model assumptions. We discuss effects, which may explain the discrepancy of our adopted values with the equivalent values:

\begin{itemize}
 \item By replacing the Pluto/Charon system with a single equivalent body, we had to assume a single average albedo for both objects, which is a coarse simplification, in order to apply our models. Furthermore, surface albedo variations of Pluto \citep{Buie2010} lead to a more complex spectral energy distribution and affect the model results significantly. Our model results would be improved by using a two-terrain model, which was used to model the thermal emission of Makemake \citep{Lim2010}, in which Pluto and Charon would each be represented by
one terrain, or applying more sophisticated albedo surface distributions: \citet{Lellouch2011} applied a 3-terrain model to model the thermal emission of Pluto and an additional terrain to model Charon's contribution.
 \item Former thermal infrared measurements of Pluto revealed its thermal lightcurve in different wavelength ranges \citep{Lellouch2000,
Lellouch2011}, which was not accounted for in our modeling. Given the time lags between the individual observations, this might have introduced a perceptible distortion of the SED, which affects the modeling results.
 \item Millimeter wavelength observations of Pluto \citep{Gurwell2005, Gurwell2010} revealed a lower surface brightness temperature than assumed from thermal equilibrium, which is due to the combination of two effects: (1) the mixing of SEDs at different temperatures, and in particular the increased contribution of the N$_2$-dominated areas, whose temperature is $\sim 37$~K due to the dominance of sublimation cooling in the thermal budget of N$_2$ ice, and (2) the fact that emissivities are expected to decrease at long wavelengths as a result of sub-surface sounding in a progressively more transparent medium. \citet{Lellouch2011} find in particular that the emissivity of CH$_4$ ice, one of the terrains in their 3-terrain Pluto model, shows a steady decrease with increasing wavelength (their Figure 8). Without additional assumptions, our thermal model is not capable of taking such effects into account. 
\end{itemize}

Due to the failed model fit and Pluto's uniqueness among the Plutinos, we exclude it from the following discussion of the properties of our Plutino sample.

\begin{figure}[t]
 \centering
  \resizebox{\hsize}{!}{\includegraphics{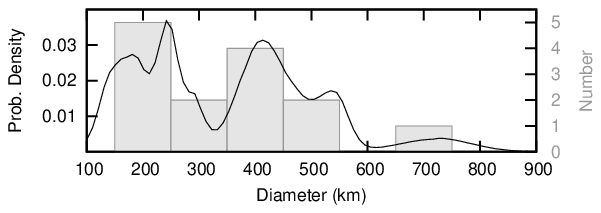}} \\
  \resizebox{\hsize}{!}{\includegraphics{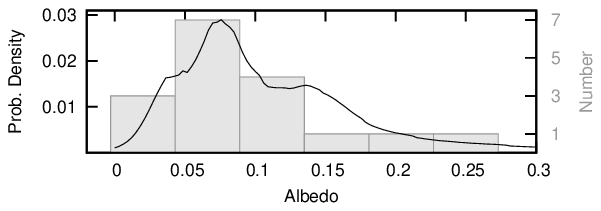}} 
 \caption{Diameter and albedo probability density functions and histograms. The probability density functions describe the probability to find an object of a specific diameter or albedo, when picking a random object. The functions were derived by summing up the probability density functions of each object, being represented by a non-symmetric normal distribution taking into account the different upper and lower uncertainties of each result. This approach returns a more detailed picture than the histogram. Pluto was excluded from the sample. }
 \label{fig:histograms} 
\end{figure}

\subsection{Sample Statistics}
\label{ref:sample_statistics}

\begin{table}[t]
 \caption{Average albedos of Centaurs, different TNO sub-populations and Jupiter Family Comets.}
 \label{tab:average_albedos}
 \centering
 \begin{tabular}{rccl}
 \hline \hline \\[-7pt]
 Population           & Average Albedo  & $N$ & Reference \\
 \hline \\ [-7pt]
 Centaurs             & $0.06$          & 21 & 1 \\
 Plutinos             & $0.08\pm0.03$   & 17 & 2 \\
 Cold Classicals$^a$  & $0.16\pm0.05$   &  6 & 3 \\
 Hot Classicals$^a$   & $0.09\pm0.05$   & 12 & 3 \\
 Scattered Disk Objects$^b$   & 0.07/0.05   & 8  & 4 \\
 Detached Objects$^b$         & 0.17/0.12$^c$  & 6 $^c$  & 4 \\
 \hline \\[-7pt]
 TNOs and Centaurs    & $0.07-0.08$     & 47 & 1 \\
 JFCs                 & $<0.075^R$      & 32 & 5  \\
 \hline
 \end{tabular}
\tablefoot{ $N$ denotes the number of datapoints.
   \tablefoottext{a}{unweighted mean;}
   \tablefoottext{b}{unweighted/weighted mean, based on relative uncertainties;}
   \tablefoottext{c}{excluding Eris;}
   \tablefoottext{R}{means the geometric albedo in $R$ band, which is comparable to $p_V$ for small values.}
   \textbf{References.} 
    (1):  \citet{Stansberry2008};
    (2):  this work;
    (3):  \citet{Vilenius2011};
    (4):  \citet{SantosSanz2011};
    (5):  \citet{Fernandez2005}.
 }
\end{table}

Figure \ref{fig:histograms} shows diameter and albedo probability density functions and histograms based on the results compiled in Table \ref{tab:adopted_results}. The size distribution ranges from 150~km to 730~km and the albedo distribution from $0.04$ to $0.28$ with a clear peak around $0.08$. The low probability density at high albedos is caused by the large uncertainties of such albedos. Excluding Pluto, the weighted mean\footnote[6]{Upper and lower uncertainties were calculated corresponding to the calculations applied in Table \ref{tab:optical_photometry}. The calculation of the weighted mean albedo based on fractional uncertainties yields 0.11. However, we prefer to adopt the weighted mean albedo based on absolute errors, since it better agrees with the low albedo of the bulk of the Plutinos and is enforced from our results in Section \ref{ref:sizedistribution}.} of the albedo, weighted by absolute uncertainties, yields $0.08\pm0.03$, which agrees well with the range of typical geometric albedos of the TNO and Centaur sample of \citet{Stansberry2008}. Table \ref{tab:average_albedos} shows a comparison of the mean albedos of our Plutino sample with different TNO sub-population and JFCs. It turns out that the average albedo of the Plutinos agrees well with that of the scattered disk, Centaur, hot classical and JFC population. Through dynamical studies \citet{Duncan1995}, \citet{Yu1999} and \citet{DiSisto2010} suggested the Plutino population to be a source of JFCs and/or Centaurs, which is compatible with our results. Furthermore, most of the Plutinos appear to have darker surfaces than the cold classical TNOs and detached objects.

\subsection{Plutino Size Distribution}
\label{ref:sizedistribution}

\begin{figure}[t]
 \centering
  \resizebox{\hsize}{!}{\includegraphics{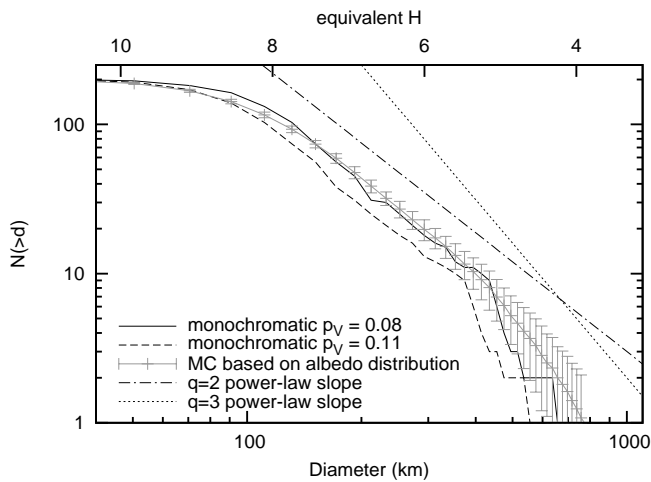}} 
 \caption{Cumulative Plutino size distributions $N(\geq d)$ based on MPC $H$ magnitudes and different assumptions: the solid and dashed line represent monochromatic albedo distributions using $p_V=0.08$ and $p_V=0.11$, respectively. The grey line shows the result of an averaged Monte Carlo simulations based on 100 runs and the measured albedo distribution with corresponding 1~$\sigma$ standard deviations. It is obvious that the monochromatic $p_V=0.08$ size distribution agrees better with the result of the Monte Carlo simulation. Furthermore, we show two cumulative power laws $N(\geq d)\sim d^{-q}$ with $q=2$ and $q=3$ and arbitrary interceptions for comparison. A $q=2$ cumulative power law better fits the measured size distribution in the size range 120--450~km, whereas the distribution of larger objects is better fit by a $q=3$ cumulative power law. We can give an upper size limit for the kink at the small-sized end around 120~km. The upper abscissa gives the equivalent $H$ magnitude based on a monochromatic albedo distribution of $p_V=0.08$. }
 \label{fig:sizedistribution} 
\end{figure}

Using the absolute magnitude estimates for all known Plutinos from the MPC and our measured albedos we are able to calibrate the Plutino size scale for the first time and to determine a cumulative size distribution of the known Plutino population. We use two different approaches to determine the cumulative size distribution $N(\geq d)$. Firstly, we assume monochromatic albedo distributions based on the two average albedos determined in Section \ref{ref:sample_statistics}. This approach enables the direct conversion of $H$ magnitudes into diameter but neglects the measured diversity in albedo. The second approach makes use of a Monte Carlo method to determine the size distribution based on the actually measured albedo distribution. For this purpose, each MPC $H$ magnitude is assigned to a randomly generated albedo, which is used to determine the diameter. Hence, we assume that $H$ is not correlated to the albedo, which is supported by our findings in Section \ref{ref:discussion_correlation}. The distribution of the randomly generated albedos follows the determined probability density function (Figure \ref{fig:histograms}). Hence, this approach takes into account the measured albedo diversity. Figure \ref{fig:sizedistribution} shows a comparison of the two different methods. It is clearly visible that the averaged Monte Carlo approach agrees well with the monochromatic $p_V=0.08$ model, which supports our assumption that this value better represents the average Plutino albedo compared to the value of $0.11$. 

For comparison, we plot in Figure \ref{fig:sizedistribution} the slopes of different cumulative power laws $N(\geq d)\sim d^{-q}$. We find that the distribution of intermediate-sized Plutinos (120-450~km) is well-described by $q=2$, which is smaller than the presumed value of $q\sim3$ \citep{Trujillo2001, Kenyon2004}. The large diameter tail ($d\geq450$~km) seems to be better described by $q=3$. However, this region suffers from uncertainties due to small number statistics. A change in the cumulative power law slope at the small-sized tail seems to occur at diameters of 80--120~km, which is larger than the proposed range of 40--80~km \citep{Bernstein2004,Kenyon2008,deElia2008} . However, the number of intrinsically fainter, and therefore smaller Plutinos is likely to be underrepresented, since the MPC sample suffers from an optical discovery bias. A proper debiasing of the size and albedo distributions is planned as part of future work. However, we suppose that a larger number of small-sized Plutinos would shift the kink to smaller diameters. Hence, we are only able to give an upper size limit of the location of the kink, which is 120~km. Furthermore, we note that our size distribution is based on albedo measurements derived from objects of the size range 150--730~km, which might be different from the albedo distribution of smaller Plutinos.

\begin{table}[t]
 \caption{Results of the modified Spearman rank correlation analysis of the Plutino sample. $N$ is the number of sample datapoints; $\langle \rho \rangle$ and $P$ are the most probable correlation coefficient and the probability of the most probable correlation coefficient to occur given no relationship in the sample (cf. text), respectively. The uncertainty of $\langle \rho \rangle$ denotes its 68~\% confidence level. $P_\sigma$ is the significance expressed in terms of $\sigma$. We show only the results of parameter pairs which meet the following criteria: $|\langle \rho \rangle| \geq 0.3$ and $P \leq 0.1$, or which are of special interest. Parameter pairs are sorted by parameter type.} 
 \label{tab:correlations}
 \centering
 \begin{tabular}{ccccc}
 \hline \hline \\[-7pt]
 Parameters & $N$ & $\langle \rho \rangle$ & $P$ & $(P_\sigma)$\\
 \hline \\ [-7pt]
 \multicolumn{5}{l}{{\bf physical parameters:}} \\
  $d$~/~$p_V$         & 17 & $-0.32_{-0.24}^{+0.29}$ & 0.217 & (1.24) \\ [+3pt]
  $d$~/~$\eta$        &  7 & $-0.12_{-0.54}^{+0.61}$ & 0.806 & (0.25) \\ [+3pt]
  $p_V$~/~$\eta$      &  7 & $0.08_{-0.67}^{+0.61}$ & 0.870 & (0.16) \\ [+3pt]
  $H$~/~$d$           & 17 & $-0.30_{-0.30}^{+0.37}$ & 0.234 & (1.19) \\ [+3pt] 
  $H$~/~$p_V$         & 17 & $-0.19_{-0.27}^{+0.30}$ & 0.468 & (0.73) \\ [+3pt] 
  $H$~/~$\eta$        &  7 & $0.08_{-0.81}^{+0.72}$ & 0.868 & (0.17) \\ [+3pt]
 \multicolumn{5}{l}{{\bf orbital and physical parameters:}} \\
  $e$~/~$d$           & 17 & $-0.62_{-0.13}^{+0.18}$ & 0.008 & (2.67) \\ [+3pt]
  $d$~/~$q$           & 17 & $0.62_{-0.17}^{+0.13}$ & 0.008 & (2.67) \\ [+3pt]
  $d$~/~$r^\star$     & 17 & $0.58_{-0.22}^{+0.16}$ & 0.015 & (2.43) \\ [+3pt]
  $p_V$~/~$r^\star$   & 17 & $0.10_{-0.32}^{+0.30}$ & 0.711 & (0.37) \\ [+3pt]
 \multicolumn{5}{l}{{\bf color information, orbital and physical parameters:}} \\
  $s$~/~$r^\star$     & 13 & $-0.58_{-0.26}^{+0.46}$ & 0.038 & (2.07) \\ [+3pt]
  $s$~/~$d$           & 13 & $-0.62_{-0.20}^{+0.32}$ & 0.025 & (2.24) \\ [+3pt]
  $s$~/~$H$           & 13 & $0.20_{-0.42}^{+0.36}$ & 0.523 & (0.64) \\ [+3pt]

  $(B-R)$~/~$d$       & 13 & $-0.50_{-0.21}^{+0.29}$ & 0.082 & (1.74) \\ [+3pt]
 \hline
 \end{tabular}
 \tablefoot{All correlation coefficients were calculated excluding Pluto. Correlation analyses taking $\eta$ into account are based solely on the 7 floating-$\eta$ fit solutions. In order to assess the effects of discovery bias, we include the heliocentric distance at the time of discovery $r^\star$ in our analysis. To improve the readability we refrain from listing all correlation values of the color indices with $r^\star$ and $d$. Instead we only show the correlation values of spectral slope with $r^\star$ and $d$, which gives similar results. }
\end{table}

\subsection{Correlations}
\label{ref:discussion_correlation}

\begin{figure*}[t]
 \centering
 \begin{tabular}{cc}
    \includegraphics[width=7.0cm]{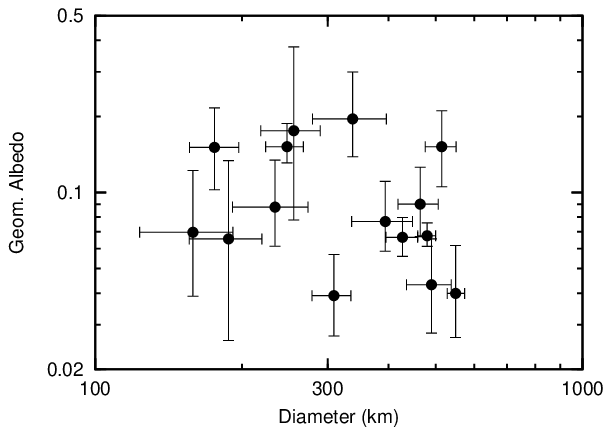} & \includegraphics[width=7.0cm]{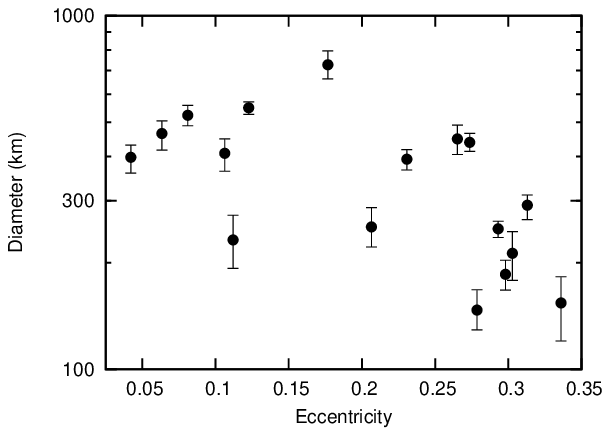} \\
    \includegraphics[width=7.0cm]{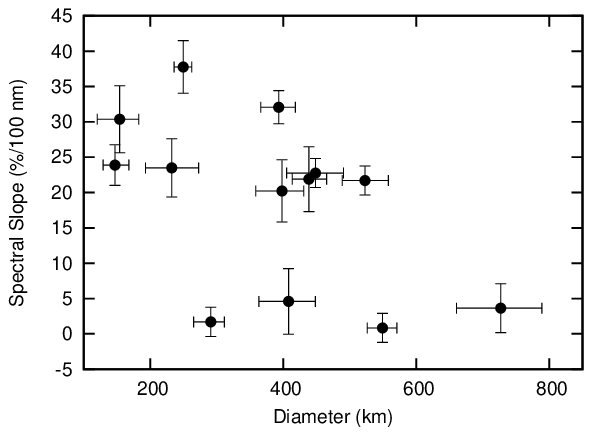} & \includegraphics[width=7.0cm]{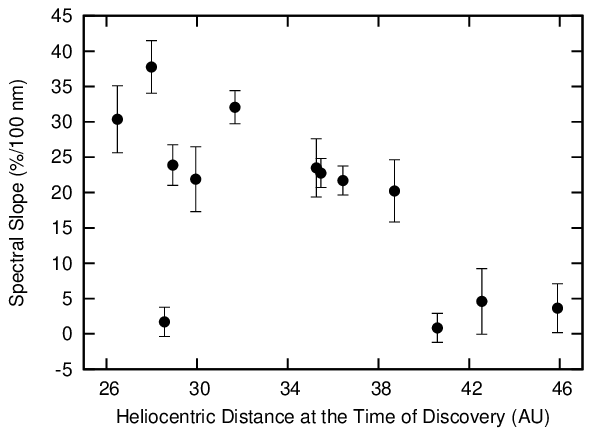} \\
 \end{tabular} 
 \caption{Plots of some correlations. {\bf Upper left}: we find no correlation between diameter and albedo; {\bf upper right}: a clear anti-correlation between eccentricity and diameter is the result of a detection bias; {\bf lower left}: we find smaller Plutinos to be redder than larger ones, which contradicts what was previously suggested by \citet{Peixinho2004} based on the intrinsical brightness $H$ of a different Plutino sample; {\bf lower right}: we find that the color of Plutinos is correlated to the heliocentric distance at the time of their discovery, which hints to a color bias.}
 \label{fig:correlation_plots} 
\end{figure*}

The Plutino sample was checked for correlations of the physical parameters, orbital properties and intrinsic optical colors, using a modified Spearman rank correlation analysis technique. The technique is based on a bootstrap method and takes into account the uncertainties (error bars) of the parameters (data points), and computes the confidence interval of the corresponding correlation coefficient. Since we are taking into account data error bars, the results are estimates of the most probable correlation coefficient $\langle \rho \rangle$ accompanied by the probability $P$ (p-value) of that $\langle \rho \rangle$ value to occur given no relationship in the sample, i.e. to occur by pure chance. The lower the value of $P$ the higher the significance level of the correlation. We present the 68~\% confidence interval for the correlation value $\langle \rho \rangle$, i.e. the interval which includes 68~\% of the bootstrapped values. This confidence interval equals the canonical 1~$\sigma$ interval of a gaussian distribution. As an additional aid in reading the $P$ values we also compute its gaussian-$\sigma$ equivalent significance $P_\sigma$. The details of the technique are described in \citet{SantosSanz2011}. Following \citet{Efron1993}, we define a weak correlation by $0.3<|\langle \rho \rangle|<0.6$ and a strong correlation by $|\langle \rho \rangle|>0.6$. $P<0.05$ indicates reasonably strong evidence of correlation, $P<0.025$ indicates strong evidence of correlation and $P<0.003$ describes a clear (3~$\sigma$) correlation.

It turns out that most parameters are uncorrelated. Only very few parameter pairs show reasonably strong evidence of correlation. Those and a few other interesting results are listed in Table \ref{tab:correlations} and discussed here. Other parameter pairs lack sufficient evidence for correlation or possess weak correlation values. 

\begin{itemize}
 \item We do not detect any strong and significant trend between diameter $d$ and albedo $p_V$ (cf. Figure \ref{fig:correlation_plots}). Although we cannot rule out the possibility of some correlation between these two parameters given our sample size, it is very unlikely that a strong correlation between the two parameters might exist. On the other hand, both diameter and albedo seem to be clearly uncorrelated with the $\eta$ value.
 \item The absolute magnitude $H$ is not correlated with diameter, albedo, or $\eta$. This shows that our sample is not biased towards large and high (or low) albedo objects, which supports our approach to randomly assign albedos to $H$ magnitudes in Section \ref{ref:sizedistribution}. This finding, however, does not exclude a discovery bias on smaller diameters than those examined here or a discovery bias based on geometrical aspects.  
 \item We find a significant anti-correlation between eccentricity $e$ and diameter $d$ (cf. Figure \ref{fig:correlation_plots}), showing that in our sample highly eccentric objects tend to be smaller. This anti-correlation is precisely the opposite of the correlation between perihelion distance $q$ and diameter $d$, and also the opposite of the correlation between diameter $d$ and the heliocentric distance of each object at the time of it's discovery $r^\star$. We consider that these relationships are merely caused by the fact that objects on more eccentric orbits come closer to the Sun, which significantly improved their detectability. Hence, the observed relation between an object's diameter and its eccentricity (and perihelion distance) is very likely to be caused by a discovery bias.
 \item The correlation between diameter $d$ and the heliocentric distance of each object at the time of its discovery $r^\star$, compared to the lack of correlation between albedo $p_V$ and $r^\star$, shows that the likelihood of detection solely depends on size, and is rather independent of the object's albedo. This follows because the brightness of an object scales as $d^2p_V$: the measured range of diameters translates into a brightness change factor of 24, whereas the total effect of the albedo can only account for a factor of 6. We can rule out a bias towards high (and low) albedos as a result of the nature of discovery of our sample targets. This supports the representativeness of the measured albedo distribution for the Plutino population and shows that the nature of the discovery bias is size-dependent.
 \item Most color indices and the spectral slope are anti-correlated with heliocentric distance at the time of an object's discovery $r^\star$ (Figure \ref{fig:correlation_plots}). The farther an object is, the less red it seems to be. This points to a probable color bias in our sample, maybe induced by the common use of $R$ band filters and the improved sensitivity in red bands of state-of-the-art detectors, which are used in TNO surveys. However, we find a lack of bluer objects at shorter heliocentric distances, which should be detectable, despite their color. This might suggest that such objects do not exist, at least within our sample.
 \item We can not confirm a trend found by \citet{HainautMBOSS} and \citet{Peixinho2004} of bluer Plutinos being intrinsically fainter using the spectral slope $s$ as a measure of color. The correlation analysis of $B-R$ and $H$ leads to similar results: $\langle \rho \rangle=0.25_{-0.43}^{+0.35}$ and $P=0.420$. Our sample suggests precisely the opposite, with a trend between spectral slope $s$ and the diameter $d$ showing smaller Plutinos to be redder (Figure \ref{fig:correlation_plots}). We note however, that \citet{Peixinho2004} find their trend due to a `cluster' of blue Plutinos with $H_R>7.5$, i.e. $H\gtrsim 8$, objects at magnitude/size ranges we do not have in our sample. Hence, both effects might be the result of selection effects. 
\end{itemize}

It is also interesting to ask if the presence of water ice on our targets (see Table \ref{tab:optical_photometry}) is correlated with albedo or diameter.  Only 6 of our
targets (other than Pluto) are known to definitely have ice, while two definitely do not. Because we do not have a quantitative and consistently defined measure of how much ice is present on these objects, a formal correlation analysis is not possible. Qualitatively speaking, we can say that 5 of the objects with ice have high albedos ($> 0.11$), with
1999~TC36 being the only icy object with a typical (0.07) Plutino albedo. Thus there is qualitative evidence that icy Plutinos have higher albedos than is typical. 1996~TP66 and 2000~GN171 are the only objects with good spectra indicating the lack of water ice. Their albedos (0.07 and 0.22, respectively) span most of the range of our measured albedos. This seems to indicate that Plutinos lacking water ice can have almost any albedo (although the lack of a detection of 70~$\mu m$ emission from 1996~TP66 by either Spitzer or Herschel casts considerable doubt on the accuracy of its albedo determination). It seems remarkable that these two objects, which are free of ices, are also the two smallest objects ($\sim 150$~km) of our sample. However, the small number of spectroscopically examined objects does not allow for a conclusion, whether the presence of ices is correlated to the object diameter. 


\section{Summary}

The analysis of the diameters and albedos of 18 Plutinos using PACS photometry, leads us to the following conclusions:

\begin{itemize}
 \item The diameter and albedo range of our Plutino sample yields $150-730$~km and $0.04-0.28$, respectively. Excluding Pluto, the weighted mean of the Plutino albedo distribution is $0.08\pm0.03$ and agrees with the average albedos observed for the scattered disk, Centaur and JFC population. This agreement is compatible with the idea that Plutinos are a source population of Centaurs and JFCs.
 \item The floating-$\eta$ fits yield a weighted mean $\eta$ for the Plutino sample of $1.11_{-0.19}^{+0.18}$, which agrees with the canonical $\eta=1.20\pm0.35$ \citep{Stansberry2008} utilized in the fixed-$\eta$ fits.
 \item We refine the size estimates of the components of 1999~TC36: $d_{A1}=272_{-19}^{+17}$~km, $d_{A2}=251_{-17}^{+16}$~km and $d_B=132_{-9}^{+8}$~km; we estimate the bulk density of the multiple system to be $\varrho=0.64_{-0.11}^{+0.15}\,\mbox{g cm}^{-3}$.  
 \item Pluto is the clear outlier of the Plutino population, being by far the largest object with the brightest surface. We have shown, that canonical simple thermal modeling of Pluto data leads to inadequate results, probably mainly due to its multi-component nature, thermal lightcurve and atmosphere.  
 \item Using our measured albedos we calibrated the Plutino size scale for the first time and determined a cumulative size distribution of the known Plutinos. We find that intermediate sized Plutinos ($120~\mbox{km} \leq D \leq 400~\mbox{km}$) follow a cumulative power-law distribution with $q=2$, whereas the distribution of larger objects is better described by $q=3$. We are able to give an upper size limit for the kink at the small-sized end of the distribution, which is 120~km.
 \item We find no correlations between albedo and diameter, as well as $H$, and diameter, albedo and $\eta$, respectively. This shows that our sample is not biased towards large and high (or low) albedo objects. Furthermore, we find a correlation between diameter and the heliocentric distance at the time of discovery, but not between the albedo and the heliocentric distance at the time of discovery. This leads us to the conclusion that the nature of the discovery bias is mainly size-dependent.
 \item A significant correlation between diameter and eccentricity (and perihelion distance) is very likely to be caused by a detectional bias based on geometrical aspects.
 \item We find hints to color biases in our sample: Plutinos, which have been farther from Sun at the time of their discovery seem to be bluer and smaller sized Plutinos tend to be redder, which contradicts previous finding by \citet{Peixinho2004}.
 \item There is qualitative evidence that icy Plutinos have higher albedos than the average of the sample. We are not able to conclude on a correlation between the diameter and the presence of ices.
\end{itemize}

\small{We would like to thank Alain Doressoundiram for his useful suggestions as referee. Furthermore, we thank Chemeda Ejeta (MPS, Katlenburg-Lindau, Germany) for performing dynamical simulations on some of our targets.
M. Mommert acknowledges support through the DFG Special Priority Program 1385: \textit{The First 10 Million Years of the Solar System - a Planetary Materials Approach}.
C. Kiss and A. Pal acknowledge the support of the Bolyai Research Fellowship of the Hungarian Academy of Sciences. P. Santos-Sanz would like to acknowledge financial support by the Centre National de la Recherche Scientifique (CNRS). J. Stansberry acknowledges support for this work provided by NASA through an award issued by JPL/Caltech. R. Duffard acknowledges financial support from the MICINN (contract Ram\'{o}n y Cajal). Part of this work was supported by the German {\it Deutsches Zentrum f\"{u}r Luft- und Raumfahrt} (DLR) project numbers 50~OR~0903, 50~OFO~0903 and 50~OR~1108, and the PECS program of the European Space Agency (ESA) and the Hungarian Space Office, PECS-98073. J.L. Ortiz acknowledges support from spanish grants AYA2008-06202-C03-01, AYA2011-30106-C02-01 and 2007-FQM2998}

\bibliographystyle{bibtex/aa}
\bibliography{bibtex/plutinos}

\end{document}